\documentclass[aps,prd,twocolumn,superscriptaddress, nofootinbib,floats,floatfix]{revtex4-1}
\usepackage{bm}
\usepackage{amsmath}
\usepackage{amsfonts}
\usepackage{aas_macros}
\usepackage{graphicx}
\usepackage[hidelinks]{hyperref}
\usepackage{booktabs}
\usepackage{color}
\usepackage[dvipsnames]{xcolor}
\usepackage[capitalise]{cleveref}
\usepackage[caption=false]{subfig}

\creflabelformat{equation}{#2\textup{#1}#3}

\DeclareMathOperator{\erf}{erf}
\DeclareMathOperator{\Ei}{Ei}
\newcommand\sun{\hbox{$\odot$}}
\newcommand\farcs{\hbox{$.\!\!^{\prime\prime}$}}
\newcommand\arcsec{\hbox{$^{\prime\prime}$}}

\begin{document}

\title{Calibrating galaxy formation effects in galactic tests of fundamental physics}

\author{D. J. Bartlett}
\email{deaglan.bartlett@physics.ox.ac.uk}
\affiliation{Astrophysics, University of Oxford, Denys Wilkinson Building, Keble Road, Oxford, OX1 3RH, UK}
\author{H. Desmond}
\affiliation{Astrophysics, University of Oxford, Denys Wilkinson Building, Keble Road, Oxford, OX1 3RH, UK}
\author{P. G. Ferreira}
\affiliation{Astrophysics, University of Oxford, Denys Wilkinson Building, Keble Road, Oxford, OX1 3RH, UK}

\begin{abstract}
    Galactic scale tests have proven to be powerful tools in constraining fundamental physics in previously under-explored regions of parameter space. The astrophysical regime which they probe is inherently complicated, and the inference methods used to make these constraints should be robust to baryonic effects. Previous analyses have assumed simple empirical models for astrophysical noise without detailed calibration or justification. We outline a framework for assessing the reliability of such methods by constructing and testing more advanced baryonic models using cosmological hydrodynamical simulations. As a case study, we use the Horizon-AGN simulation to investigate warping of stellar disks and offsets between gas and stars within galaxies, which are powerful probes of screened fifth forces. We show that the degree of `U'-shaped warping of galaxies is well modelled by Gaussian random noise, but that the magnitude of the gas--star offset is correlated with the virial radius of the host halo. By incorporating this correlation we confirm recent results ruling out astrophysically relevant Hu-Sawicki $f(R)$ gravity, and identify a $\sim30\%$ systematic uncertainty due to baryonic physics.
    Such an analysis must be performed case-by-case for future galactic tests of fundamental physics.
\end{abstract}

\maketitle

\section{Introduction}
\label{sec:Introduction}

Tensions between different probes of the Universe \cite{Cosmology_Intertwined_1,Cosmology_Intertwined_2,Cosmology_Intertwined_3} as well as small-scale observational inconsistencies \cite{Bullock, Del_Popolo_2017}, coupled with the cosmological constant problem \cite{Padilla_2015}, lead us to question whether General Relativity (GR) is the correct description of gravity. 

Traditional tests probe gravity in one of three regimes: small scales (laboratory or Solar System, e.g. lunar laser ranging \cite{Nordtvedt_1968,Murphy_2012}), cosmological scales (e.g. cosmic microwave background \cite{Planck_VI_2018}), or the strong-field regime (e.g. gravitational waves from binary black holes \cite{LIGO_GR_Tests_2020}). Notably absent is the regime occupied by galaxies where very few tests have been conducted  \cite{Baker_2015}. The field of astrophysical tests of gravity is nonetheless emerging as powerful and complementary to traditional methods \cite{Novel_Probes_2019}.

There are clear advantages to studying galaxies. We are no longer restricted to the linear regime, so remove the associated limits on our results introduced by cosmic variance. However, the same non-linearities that make galaxies such rich laboratories for investigating fundamental physics also complicate any such analysis, because the complex astrophysical effects that shape galaxies act as critical systematics.

Bayesian Monte Carlo-based forward models have proven to be successful in constraining fundamental physics on galactic scales \cite{Desmond_2018,Desmond_2018_warp,Desmond_letter,Pardo_2019,f(R)_ruled_out,VS_BH_constraints}. These analyses have however assumed simple empirical noise models in which astrophysical contributions to the signals are assumed to be Gaussian distributed and uncorrelated with the properties of galaxies and their environments. The inferences would be biased if astrophysical effects were in fact significantly degenerate with the fundamental physics being tested. In this work we propose an approach for constructing reliable noise models based on correlations found in cosmological hydrodynamical simulations between relevant parameters of the system.

As a case study, we consider warping of stellar disks and offsets between the stellar and gas mass centroids of galaxies. These are important probes of screened fifth forces \cite{Jain_Vanderplas, Vikram_2013, Desmond_2018, Desmond_2018_warp}, and have recently \cite{f(R)_ruled_out} been used to rule out astrophysically relevant Hu-Sawicki $f(R)$ gravity \cite{Hu_2007}, a paradigmatic modified gravity model. A galactic disk can be warped through a plethora of physical phenomena besides modified gravity, including gas infall into the dark matter halo \cite{Ostriker_1989}, dark matter self interactions \cite{Secco, Pardo_2019}, or interaction of the disk with companions \cite{Weinberg_1998,Semczuk_2020} (see \cite{Binney_1992} for a review). Furthermore, gas can be displaced from the centres of galaxies in clusters due to a combination of ram pressure stripping and tidal interactions \cite{Scott_2010}. It is therefore likely that disks will be warped and the stellar and gas mass separated even in the absence of a fifth force. This was accounted for in Refs.~\cite{Desmond_2018, Desmond_2018_warp, f(R)_ruled_out} by convolving the fifth force likelihood with a Gaussian noise model with a width that was either constant between galaxies or proportional to their distance. It is however unclear that this model should be sufficiently flexible to account for baryonic physics, which will make the noise a function of galaxies' properties and environments.

In \Cref{sec:Theory} we introduce our case study and in \Cref{sec:General method} we outline the criteria used to evaluate the suitability of the astrophysical noise model. The cosmological hydrodynamical simulation in $\Lambda$CDM that we will use to assess this model, Horizon-AGN, is described in \Cref{sec:Horizon-AGN}.
We detail our measurement and modelling of the signals in the simulation in \Cref{sec:Methods} and verify that we obtain a null detection of a fifth force in \Cref{sec:Constraints}. We investigate the validity of the Gaussian noise model used to make these constraints in \Cref{sec:Validity of noise model}, and discuss the impact of the assumed halo density profile for this model in \Cref{sec:Validity of halo density model}. We discuss the broader context of our work, and conclude, in \Cref{sec:Conclusions}.

\section{Case study --- screened fifth forces}
\label{sec:Theory}

While our methodology will prove to be general, we choose a specific case study to show it working in practice. We focus on aspects of galaxy morphology (gas--star offsets and warping of stellar disks) caused by thin-shell-screened fifth forces generated by a new light scalar gravitational degree of freedom. Throughout the paper we use units in which $\hbar \equiv c \equiv 1$.

\subsection{Theoretical background}

Scalar-tensor theories introduce at least one additional scalar, $\phi$, which can couple to matter and is typically sourced via a generalised Poisson equation,
\begin{equation}
\label{eq:generalised Poisson}
    Z^{ij} \partial_i \partial_j \phi + m^2 \phi = 8 \pi G_{\rm N} \alpha \rho,
\end{equation}
where $\rho$ is the matter density, $G_{\rm N}$ is Newton's constant, $Z^{ij}$ is a generalised kinetic coefficient, $m$ is the mass of the scalar and $\alpha$ the strength of its coupling to matter. These parameters are obtained from the full equation of motion of $\phi$ by expanding about a background value, $\phi_0$, and applying the quasi-static limit. This scalar then produces an additional acceleration
\begin{equation}
    \bm{a}_5 = - \alpha \nabla \phi,
\end{equation}
(the fifth force), which leads to an effective enhancement of the gravitational force,
\begin{equation}
    G_{\rm N} \to G_{\rm N} \left( 1 + \frac{\Delta G}{G_{\rm N}} \right),
\end{equation}
where $\Delta G/G_{\rm N} \equiv 2 \alpha^2$.
In the absence of a screening mechanism, the parameters $\alpha$ and $m$ must be tuned to very small values in order to obey Solar System and laboratory tests of gravity \cite{Burrage_2018} (see \cite{Jain_2010,Joyce_2015,Khoury_2010,Novel_Probes_2019} for reviews of screened modified gravity theories). Screening dynamically suppresses the kinetic, mass or coupling terms in \autoref{eq:generalised Poisson} by allowing them to depend on the background scalar field, resulting in either `kinetic' (e.g. K-mouflage \cite{Babichev_2009} and Vainshtein \cite{Vainshtein_1972}) or `thin-shell' (e.g. chameleon \cite{Khoury_2004a,Khoury_2004b}, symmetron \cite{Hinterbichler_2010} and dilaton \cite{Brax_2010}) screening. An alternative mechanism utilises dark matter-baryon interactions to produce a dark matter density-dependent gravitational constant \cite{Sakstein_2019}.

In this work we focus on thin-shell screening mechanisms ($Z^{ij} = \delta^{ij}$), where the degree of suppression is determined by the gravitational potential, $\Phi$, such that an object is approximately unscreened if $|\Phi| < \chi$ and screened otherwise, where $\chi$ is the theory-dependent ``self-screening parameter''. The archetypal example is $f(R)$ gravity \cite{Buchdahl_1970,Carroll_2004}, which is obtained from the Einstein-Hilbert action of GR by replacing the Ricci scalar, $R$, with $R + f(R)$, i.e.
\begin{equation}
    S = \int {\rm d}^4 x \sqrt{-g} \: \frac{R + f(R)}{16 \pi G_{\rm N}} + S_{\rm m},
\end{equation}
for matter action $S_{\rm m}$ (for reviews of $f(R)$ gravity see \cite{Sotiriou_2010,DeFelice_2010}). The propagating degree of freedom of $f(R)$ gravity is $f_{R} \equiv \frac{{\rm d}f}{{\rm d}R}$, with a background value today of $f_{R 0}$. We will phrase our results in terms of the Compton wavelength, $\lambda_{\rm c}$, of the scalar field (applicable to any scalar-tensor theory), or equivalently $f_{R0}$ in the Hu-Sawicki model of $f(R)$.
Astrophysical tests are relatively insensitive to the specific theory \cite{Sakstein_2015} and our results are applicable to all thin-shell-screened theories with an astrophysical range fifth force. We will compute $\Phi$ sourced by matter within $\lambda_{\rm c}$ of an object \cite{Cabre_2012,Zhao_2011a,Zhao_2011b}, and use the screening cutoff
\begin{equation}
\label{eq:f(R) critical potential}
    \chi = \frac{3}{2} f_{R0} = \frac{3}{2} \times 10^{-8} \left( \frac{\lambda_{\rm c}}{0.32 {\rm \, Mpc}} \right)^2,
\end{equation}
appropriate for the $n=1$ Hu-Sawicki model \cite{Hu_2007}.

\subsection{Observables: gas-star offsets and galaxy warps}

Main sequence stars will always be screened if $\chi \lesssim 10^{-6}$, since this is approximately the Newtonian potential at their surfaces. On the other hand, if a galaxy is in a sufficiently low density environment, the gas and dark matter within the galaxy can be unscreened. Therefore different components of a galaxy can experience different accelerations, and thus the Equivalence Principle is violated. 

\begin{figure}
	 \includegraphics[width=\columnwidth]{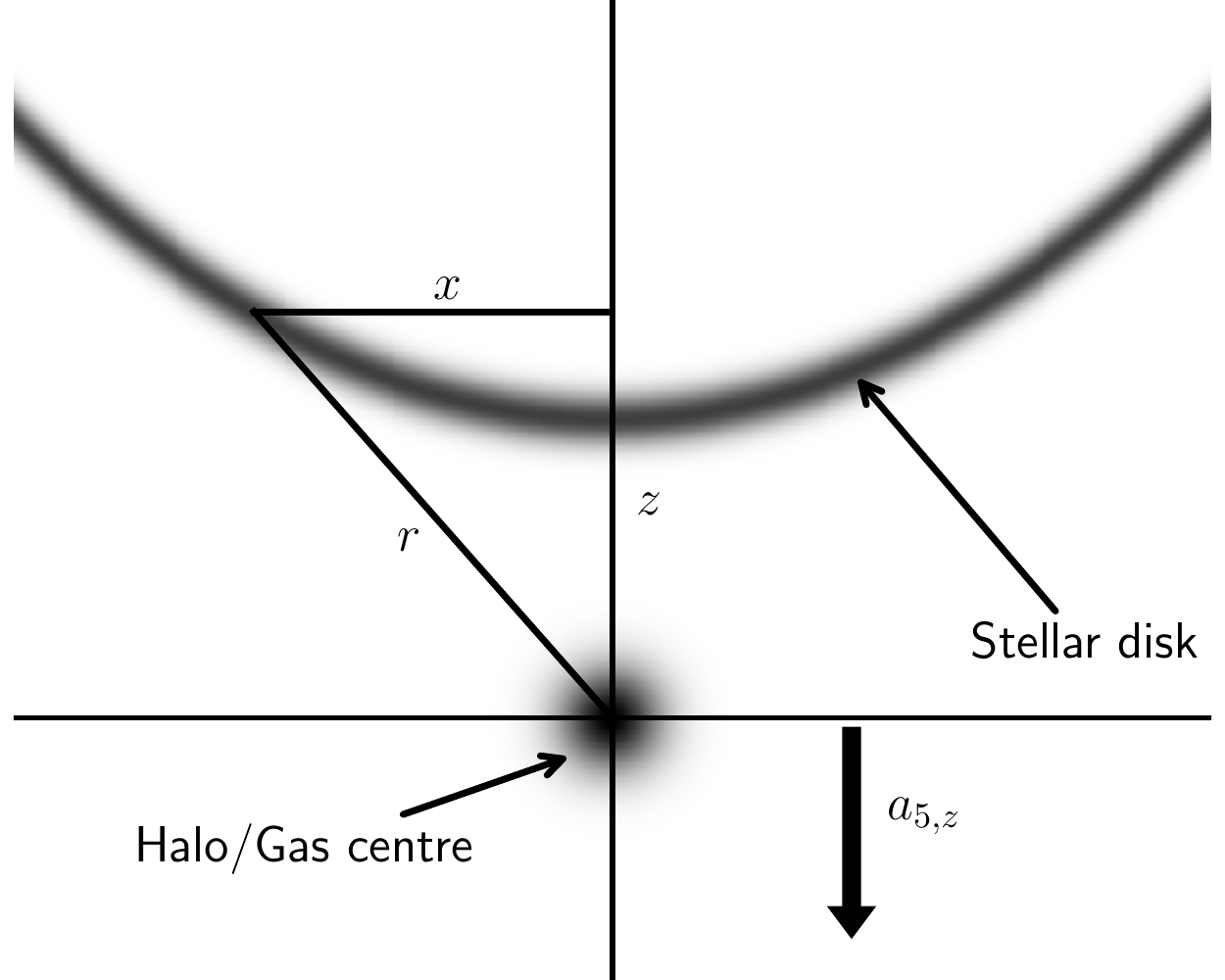}
	 \caption{\label{fig:warp_diagram}Schematic diagram of the displacement between the gas and star centres and the formation of a warped stellar disk due to a thin-shell screened fifth force, $a_{5}$. Unlike the gas and dark matter, the stars typically do not feel the fifth force since they are self-screened, resulting in these two morphological features.}
\end{figure}

This generates two key morphological signals, as illustrated in \autoref{fig:warp_diagram}. The first is that the centre of the galaxy as measured by the gas will not coincide with the centre as measured by the stars. For an external fifth force field $\bm{a}_5$, evaluated with $\Delta G / G_{\rm N} = 1$, the displacement of the gas centre from the stellar, $\bm{r}_{\star} = r_{\star} \hat{r}_{\star}$, is
\begin{equation}
    \label{eq:General offset}
    \frac{G_{\rm N} M \left( r_{\star} \right)}{r_{\star}^2} \hat{r}_{\star} = \frac{\Delta G}{G_{\rm N}} \bm{a}_5,
\end{equation}
where $M(r)$ is the enclosed mass at a distance $r$ from the halo centre.

This displacement results in a gravitational potential gradient
across the stellar disk, which warps the disk in a characteristic `U' shape. 
For equilibrium, we require the total acceleration to be constant along the disk. By equating the gravitational and fifth force contributions, we
find the displacement, $z$, normal to the the major axis of the disk, $x$, in the plane of the sky to be
\cite{Desmond_2018_warp}
\begin{equation}
    \label{eq:General z}
    z \left ( x \right) = 
    - \frac{\Delta G}{G_{\rm N}^2} \frac{|r|^3}{M \left( r \right)} \bm{a}_5 \cdot \hat{z}
    \approx - \frac{\Delta G}{G_{\rm N}^2} \frac{|x|^3}{M \left( x \right)} \bm{a}_5 \cdot \hat{z},
\end{equation}
where we approximate $z \ll x$ for the second equality. Note that near the centre of the disk this approximation does not hold, so $z\left(x=0\right)$ is not necessarily zero, although this does not affect $w_1$ below.
The disk therefore bends in the opposite direction to the projection of $\bm{a}_5$ onto the disk normal.
The magnitude of this warp can be described by the warp statistic
\begin{equation}
    \label{eq:General w1}
    w_1 \equiv \frac{1}{L^3} \int_{-L}^{L} |x| \left( z \left( x \right) - \left< z \right> \right) {\rm d} x,
\end{equation}
where we choose $L = 3 R_{\rm eff}$, as in \cite{Desmond_2018_warp,f(R)_ruled_out}. The modulus sign in the integral picks out specifically `U'-shaped warps, as opposed to the more commonly observed `S'-shaped warps \cite{Bosma_1991,GarciaRuiz_2002}.

\section{Assessing the impact of baryons}
\label{sec:General method}

To infer $\Delta G / G_{\rm N}$ as a function of $\lambda_{\rm c}$, we construct a galaxy-by-galaxy Bayesian forward model. This consists of two steps. First, we model our target observable (signal; $\bm{r}_\star$ or $w_1$) as a function of our new physics parameters ($\Delta G / G_{\rm N}$ and $\lambda_{\rm c}$) and the properties of the galaxy considered (e.g. the halo density profile) by evaluating \autoref{eq:General offset} or \autoref{eq:General w1}. Any uncertainty in these properties would turn the predicted signal into a distribution, the likelihood function. We combine this with the second part of the model that describes other processes (noise) that could lead to the same observable and hence alter the prediction due to new physics. We then constrain the new physics parameters and the noise together by comparing to observations with a Markov Chain Monte Carlo algorithm.

Once we have specified our model, we must check for systematic uncertainties which could bias the inference. The use of cosmological hydrodynamical simulations for this purpose offers two advantages: i) we know exactly what the theoretical parameters and implementation of baryonic physics are in the simulation, and ii) we have more information available there than we do observationally.

In particular, there are three questions to consider:
\begin{enumerate}
    \item Are there any correlations between galactic properties and the target observable in the simulation that are not accounted for in the noise model?
    \item How significant are those correlations in the inference?
    \item Is the model sufficient in light of the extra information available in the simulation?
\end{enumerate}

To answer point 1, we investigate whether we can predict the simulated observable from the parameters used to predict the signal in the context of the new physics model. Investigating potentially complex correlations without an a priori known functional form is best done in a machine-learning context, using algorithms to adaptively determine the functions to employ. For our case study, we train a Random Forest regressor on the simulated data; by fitting nonlinear decision trees to predict the signal from other variables, the regressor is able to assign a relative importance to each feature for determining the simulated signal \citep{scikit-learn}. Of course, to ensure our conclusions are robust to the choice of estimator, one should try multiple approaches. For our example, we repeat the analysis using an Extra Trees regressor and obtain consistent results.

If no significant correlations are found (and the real universe is similar to the simulated one), it is justified to model baryonic noise through uncorrelated random variables. Conversely, if one or more parameter is found to correlate with the simulated signal, then this model may not be sufficient. To quantify this, one may then compare the constraints obtained from the simulated data using different noise models. If a simple parameterisation can be found between the simulated signal and galactic properties due to baryonic physics, then one could allow the parameters of the noise model to vary continuously with these properties and hence marginalise over them. Alternatively, one could use one or more galactic properties to sort the sample into bins, and fit a separate noise model within each bin. 
If the difference in the constraint between these two methods is within some specified tolerance, then one can conclude that the simplified model is adequate; otherwise one should use the more complex one.

The extra data afforded by a simulation may be used to check other aspects of the inference method as well, for example unobservable properties of galaxies' halos. In data these must be modelled using observables that are available, which can introduce significant uncertainty. For the simulated data, however, one can compare the results of using the ``true'' vs model parameters to assess the accuracy of the model. In our example we consider the inner power law slope of the halo density profile: in the observational sample of \cite{f(R)_ruled_out} the absence of dynamical information at small radius makes this unobservable (it is estimated using halo abundance matching), but it can be measured directly from the dark matter particles in the simulation.

\section{The Horizon-AGN simulation}
\label{sec:Horizon-AGN}

We explore the fifth force inferences of Refs.~\cite{Desmond_2018, Desmond_2018_warp, f(R)_ruled_out} in the context of Horizon-AGN, a $(100 {\rm \, Mpc}/h)^3$ cosmological hydrodynamical simulation\footnote{\url{http://www.horizon-simulation.org/about.html}} \citep{Dubois_2014}. The simulation was run with the Adaptive Mesh Refinement code \textsc{ramses} \citep{Teyssier_2002}. The maximum refinement gives an effective physical resolution of $\Delta x = 1 {\rm \, kpc}$, where a new refinement level is added whenever the mass in that cell exceeds 8 times the initial mass resolution. The force softening scale is $\sim 2 {\rm \, kpc}$.

The WMAP-7 cosmology \cite{Komatsu_2011} is adopted, so we consider a $\Lambda$CDM universe with total matter density $\Omega_{\rm m} = 0.272$, dark energy density $\Omega_{\rm \Lambda} = 0.728$, amplitude of the matter power spectrum $\sigma_8 = 0.81$, baryon density $\Omega_{\rm b} = 0.045$, Hubble constant $H_0 = 70.4 {\rm \, km \, s^{-1} \, Mpc^{-1}}$, and power spectrum slope $n_{\rm s} = 0.967$. The simulation contains $1024^3$ dark matter particles, giving a dark matter mass resolution of $M_{\rm DM, \, res} = 8 \times 10^7 {\rm \, M_{\sun}}$.

Importantly, several baryonic effects are accounted for, including prescriptions for background UV heating, gas cooling, and feedback from stellar winds and type Ia and type II supernovae assuming a Salpeter initial mass function (IMF) \cite{Dubois_2008,taysun2012a}. Stars are formed with a density threshold of $n_0 = 0.1 {\rm \, H \, cm^{-3}}$ using a Schmidt law with 1 per cent efficiency \cite{Rasera_2006}, with a stellar mass resolution of $M_{\rm \star, \, res} = 2 \times 10^6 {\rm \, M_{\sun}}$.

The \textsc{adaptaHOP} structure finder \citep{Aubert_2004,Tweed_2009} is used to identify halos and galaxies from the dark matter and star particles, respectively. The smoothed density field obtained from the 20 nearest neighbours must exceed 178 times the mean total matter density \cite{Gunn_1972}, and a minimum of 50 particles is required to define a structure. We obtain the centre of the halo or galaxy by applying a shrinking sphere approach \citep{Power_2003} to find the position of the densest particle; the halo (galaxy) centre is at the position of the densest dark matter (star) particle.

As in \cite{Chisari_2017,Bartlett_2020}, we produce galaxy+halo structures by matching the most massive unassigned galaxy to a halo, provided its centre is within 10 per cent of the virial radius, $r_{\rm vir}$, of the halo. Each halo is considered in turn, moving from the most to least massive. Of the initial 126,361 galaxies identified, 117,099 are partnered with a halo with this procedure.

\section{Methods}
\label{sec:Methods}

\subsection{Measuring the offsets and warps}

\subsubsection{Stellar warp}
\label{sec:Stellar warp}

For every galaxy, ${\rm g}$, with its centre at $\bm{r}_{\rm g}$ relative to the centre of the simulation volume, we find all star particles identified by the galaxy finder as belonging to that galaxy. We take the coordinates of each star particle, $\bm{r}_i$, and project these into a plane containing the angular momentum of the galaxy, $\bm{J}_{\rm g}$, to increase the probability of viewing a disk edge on. To make this projection, we define two orthogonal unit vectors for each galaxy
\begin{equation}
	\hat{e}_1 \equiv \frac{\bm{J}_{\rm g}}{\left| \bm{J}_{\rm g} \right|}, \qquad \hat{e}_2 \equiv \frac{\bm{r}_{\rm g} \times \bm{J}_{\rm g}}{\left| \bm{r}_{\rm g} \times \bm{J}_{\rm g} \right|},
\end{equation}
and hence find the projected (angular) coordinates of the $i^{\rm th}$ star particle to be
\begin{equation}
	\tilde{x}_i \equiv \frac{\bm{r}_i \cdot \hat{e}_1}{\left| \bm{r}_{\rm g} \right|}, \qquad \tilde{z}_i \equiv  \frac{\bm{r}_i \cdot \hat{e}_2}{\left| \bm{r}_{\rm g} \right|}.
\end{equation}

We fit the distribution of these particles to a S\'{e}rsic \citep{Sersic_1963} profile of index $n$, such that the probability of having a particle at $(x_i, z_i) = (\tilde{x}_i - x_0, \tilde{z}_i - z_0)$ is
\begin{equation}
	p \left( x_i, z_i \right) = I_0 \exp \left( - b_n \left[ \left( \frac{R_i}{R_{\rm eff}} \right)^{\frac{1}{n}} - 1 \right] \right),
\end{equation}
where $R_i$ is the two-dimensional distance from the centre of the distribution. The major axis of the elliptical contours, of ellipticity $\epsilon \in [0, 1)$, is at an angle $\theta$ relative to the $x$ axis,
where
\begin{equation}
    \epsilon \equiv 1 - \frac{b}{a},
\end{equation}
for major and minor axis lengths $a$ and $b$ respectively.
The normalisation constant, $I_0$, is
\begin{equation}
	I_0 = \frac{b_n^{2n} e^{-b_n}}{2 \pi n R_{\rm eff}^2 \Gamma \left( 2n \right) \left( 1 - \epsilon \right)},
\end{equation}
where we have defined $R_{\rm eff}$ such that half of the probability lies within $R_{\rm eff}$, 
\begin{equation}
	\frac{\gamma \left( b_n, 2n \right)}{\Gamma \left( 2n \right)} = \frac{1}{2},
\end{equation}
where $\gamma$ is the incomplete lower gamma function. We fit for the 6 parameters of this distribution ($ R_{\rm eff}, n,  x_0,  z_0,  \epsilon, \theta$) by maximising the likelihood
\begin{equation}
	\log \mathcal{L}_{\rm g} \left( R_{\rm eff}, n,  x_0,  z_0,  \epsilon, \theta \right) = \sum_{i \in {\rm g}} \log p \left( x_i, z_i \right).
\end{equation}

We enforce uniform priors on all parameters in the ranges given in \autoref{tab:Sersic priors}, so that the maximum likelihood is also the maximum of the posterior. The optimisation is run five times for each galaxy using Powell's method \cite{Powell_1964}, with a different randomly generated start point each time. We adopt the maximum likelihood of these five as the true maximum likelihood, but require that at least two other converged points have parameters within 5 per cent of the maximum likelihood point. Otherwise, we say that the fit has not converged and we run the fit five more times. Once again we find the maximum likelihood point (of the ten) and require two different points to have parameters within 5 per cent of it. We keep adding five more fits until we obtain a converged result. After repeating this procedure five times, we find that we have successfully fitted over 99 per cent of the galaxies.

After six runs, 9 of our galaxies have zero likelihood for every iteration of the fit (i.e. this has been returned 30 times). Upon inspection of these galaxies, we find that they contain fewer than 75 star particles. It is not surprising, therefore, that we cannot fit a distribution to them. The shape measurements are unlikely to be reliable if we have too few particles, and we therefore reject galaxies with masses below $2 \times 10^9 {\rm \, M_{\sun}}$. Given that the dark matter resolution is $40$ times coarser than the stellar mass resolution (\Cref{sec:Horizon-AGN}), we implement a corresponding minimum halo mass of $8 \times 10^{10} {\rm \, M_{\sun}}$. Both of these cuts are also implemented when generating the sample for the gas--star offset inference. We find that changing these mass cuts by $\pm 50$ per cent does not significantly affect our results.

\begin{table}
    \caption{Priors used in the S\'{e}rsic fit to the star particles for each galaxy, where the symbols are defined in \Cref{sec:Stellar warp}. All priors are uniform in the range given.}
    \label{tab:Sersic priors}
    \centering
    \begin{tabular}{l|l|c}
    \textbf{Parameter} & \textbf{Description} & \textbf{Prior} \\
    \hline
    $R_{\rm eff}$ & Effective radius. & $> 0$\\
    $n$ & S\'{e}rsic index. & $> 0$\\
    $x_0$ & Centre of profile along $x$. & - \\
    $z_0$ & Centre of profile along $z$. & - \\
    $\epsilon$ & Ellipticity. & $[0, 1)$\\
    $\theta$ & Angle between major axis and $x$. & $[0, \pi)$ \\
    \end{tabular}
\end{table}

The observational warp statistic is determined from an image, so we must generate mock images of galaxies from Horizon-AGN. Since the optical data in \cite{Desmond_2018_warp,f(R)_ruled_out} is from the Nasa Sloan Atlas (NSA)\footnote{\url{ https://www.sdss.org/dr13/manga/manga-target-selection/nsa/}}, we use the pixel size for \textit{i-} and \textit{r-}band images from the Sloan Digital Sky Survey (SDSS) \cite{SDSS_DR13} ($\Delta x = \Delta z = 0\farcs 396$) as the NSA predominately contains sources from SDSS. We use the projected coordinates of all star particles within a galaxy in the $\hat{e}_1-\hat{e}_2$ plane to determine the intensity map $I(x, z)$. From this we find the luminosity-weighted $z$ position as a function of $x$,
\begin{equation}
\label{eq:zbar definition}
	\bar{z} \left( x \right) = \frac{\sum_{z = - L_z}^{L_z} z I (x,z) }{\sum_{z = - L_z}^{L_z} I(x,z)},
\end{equation}
where $L_z = N_{\rm eff} (b/a)R_{\rm eff}$, and the sum is over all pixels, which have $x$ and $z$ spacing $\Delta x$ and $\Delta z$ respectively. As in \cite{Desmond_2018_warp,f(R)_ruled_out}, we choose $N_{\rm eff} = 3$. We calculate the mean of $\bar{z}$ across the whole image
\begin{equation}
	\left< z \right> = \frac{1}{n_x} \sum_{x = - N_{\rm eff} R_{\rm eff}}^{N_{\rm eff} R_{\rm eff}} \bar{z} (x),
\end{equation}
where we have $n_x$ grid points along the $x$ axis, and we subtract this from $\bar{z}$ to have a variable of zero mean,
\begin{equation}
	\bar{z}^\prime (x) \equiv \bar{z}(x) - \left< z \right>.
\end{equation}
Finally, we use this variable to calculate the warp statistic
\begin{equation}
	w_1 = \frac{1}{ \left( N_{\rm eff} R_{\rm eff} \right)^3}  \sum_{x = - N_{\rm eff} R_{\rm eff}}^{N_{\rm eff} R_{\rm eff}} |x| \bar{z}^\prime (x) \Delta x.
\end{equation}

Just creating a 2D histogram of the star particles onto the grid described above gives too many columns for which the intensity is zero, and thus \autoref{eq:zbar definition} is undefined. This is not a problem in the SDSS images where such zero-intensity columns are not found. To circumvent this problem we smear each point-like star particles into a Gaussian with a standard deviation equal to the pixel width. This procedure works well for the majority of galaxies, and we discard those for which we still have columns of zero intensity.

\subsubsection{Gas-star offset}
\label{sec:Gas-star offset}

\begin{figure}
	 \includegraphics[width=\columnwidth]{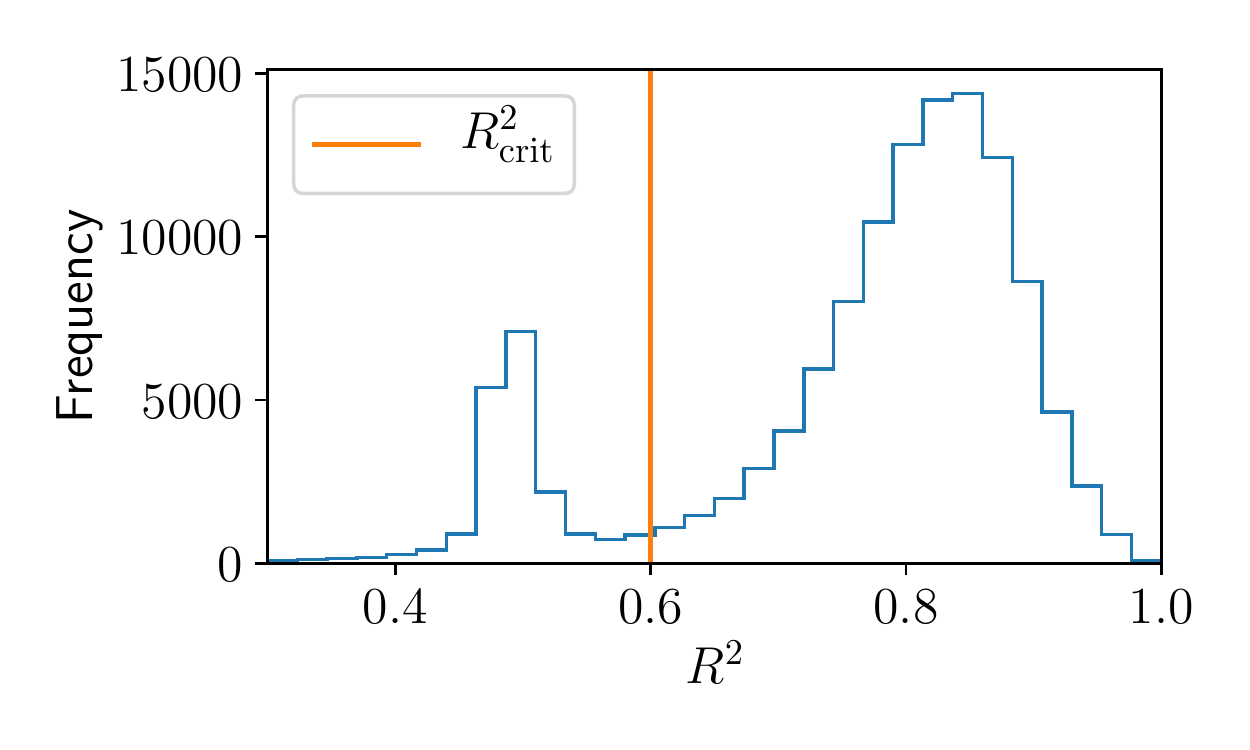}
	 \caption{\label{fig:gas_R2_distribution} The distribution in $R^2$ values (\autoref{eq:R2 definition}) for the fits to the gas density surrounding a galaxy to a Gaussian. The distribution is bimodal, indicating that a cut of $R^2_{\rm crit} = 0.6$ is sufficient to remove poor fits.}
\end{figure}

The gas centre of the galaxy is obtained by considering all gas within a box centred on the position of the densest star particle, extending to $\pm N_{\star} R_{\rm eff}$ in each dimension, where we choose $N_{\rm \star} = 4$. If we simply calculated the centre of mass of the gas in this box, we would bias our results towards small offsets; the extreme case of a uniform gas density distribution would have a centre of mass at the origin (so zero offset) even though there is no physical justification for this. Instead, we fit the density profile to a three dimensional Gaussian,
\begin{equation}
    \rho_{\rm gas} \left( \bm{r} \right) = \rho_{\rm g} \exp \left( - \frac{1}{2} \left(\bm{r} - \bm{\mu}\right)^{\rm T} \Sigma_{\rm g}^{-1} \left(\bm{r} - \bm{\mu}\right) \right),
\end{equation}
where we fit for the central density, $\rho_{\rm g}$, mean position, $\bm{\mu}$, and the covariance matrix, $\Sigma_{\rm g}$. The fitted mean is then taken to be the gas centre, and we project the resulting offset between this and the centre of mass of the star particles into right ascension (RA), $r_{\star,\alpha}$, and declination (Dec), $r_{\star,\delta}$, components. To determine convergence, we calculate the $R^2$ value for the fit, 
\begin{equation}
    \label{eq:R2 definition}
    R^2 \equiv 1 - \frac{\sum \left( \rho_{\rm meas} \left( \bm{r} \right) - \rho_{\rm gas} \left( \bm{r} \right) \right)^2}{\sum \left( \rho_{\rm meas} \left( \bm{r} \right) - \bar{\rho} \right)^2}
\end{equation}
where $\rho_{\rm meas}$ is the measured gas density, $\bar{\rho}$ is the mean of $\rho_{\rm meas}$, and the sum runs over all cells within the box of gas considered. We plot these in \autoref{fig:gas_R2_distribution} and
see that the distribution is bimodal, suggesting that a cut in $R^2$ of $R^2_{\rm crit} = 0.6$ is sufficient to remove the poorly fitted density fields. We have repeated the analysis with $R^2_{\rm crit}$ in the range $0.5-0.8$ and find that the constraint is relatively insensitive to this parameter.

To ensure that the gas is associated with the galaxy of interest, we define a characteristic length scale of the gas
\begin{equation}
    l_{\rm g} \equiv \left( \det \Sigma_{\rm g} \right)^{\frac{1}{6}},
\end{equation}
which is the geometric mean of the standard deviations of the density distribution along the principal axes. We then remove all galaxies where the gas--star offset is larger than $N_{\rm g} l_{\rm g}$, where we choose $N_{\rm g} = 4$. Varying $N_{\rm g}$ in the range $2-5$ changes the constraint by less than 50 per cent, so this choice is not important. Further, to prevent the gas associated with nearby galaxies from affecting our results, we remove all galaxies from our sample whose nearest neighbour is within $N_{\rm nn} = N_{\rm \star}$ times the sum of the effective radii of the galaxies.

\subsection{Halo density profile}
\label{sec:Halo density profile}

As in \cite{f(R)_ruled_out,Desmond_2018,Desmond_2018_warp}, in \Cref{sec:Modelling the halo restoring force} we will forward model the gas--star offsets and galaxy warps assuming that the halo density profile is a power law within some transition radius, $r_{\rm t}$,
\begin{equation}
\label{eq:Power law density}
    \rho \left( r \right) = \rho_{\rm t} \left( \frac{r}{r_{\rm t}} \right)^{- \beta}.
\end{equation}
For each galaxy, we therefore need to know $r_{\rm t}$, the density at $r_{\rm t}$ ($\rho_{\rm t}$), and the inner power law slope ($\beta$). In \cite{Desmond_2018}, abundance matching (AM) is used to find the Navarro-Frenk-White (NFW) \cite{Navarro_1996} profile parameters for the host halo of each galaxy,
\begin{equation}
\label{eq:nfw_profile}
	\rho_{\rm NFW} \left( r \right) = \frac{\rho_0}{\left( r / r_{\rm s} \right) \left( 1 + r / r_{\rm s} \right)^{2}}.
\end{equation}
It is then assumed that $r_{\rm t} = r_{\rm s}$ and that $\beta = 0.5$.

As discussed in \Cref{sec:Horizon-AGN}, the galaxies in Horizon-AGN are already matched to halos, but we do need to fit for the NFW parameters. To do this, we use that the probability of some particle to be at radius $r$ is
\begin{equation}
	P \left( r | r < r_{\rm max} \right) = \frac{4 \pi r^2 \rho \left( r \right)}{M \left( r_{\rm max} \right)},
\end{equation}
where 
\begin{equation}
	M \left( r_{\rm max} \right) = \int_0^{r_{\rm max}} 4 \pi r^2 \rho \left( r \right) {\rm d}r. 
\end{equation}
From this, it is clear than any dependence on $\rho_0$ in $P \left( r | r < r_{\rm max} \right)$ cancels, so we must determine this at the end. To fit the profile, we maximise
\begin{equation}
	\log \mathcal{L}_{\rm h} = \sum_{i} \log \left( P \left( r_i | r_i < r_{\rm max} \right) \right),
\end{equation}
where the sum is over all dark matter particles identified as belonging to the halo within some radius $r_{\rm max}$ of the halo centre of mass. We therefore fit for $r_{\rm s}$, requiring that $r_{\rm s} > 0$. To find the second parameter of this profile ($\rho_0$), we enforce
\begin{equation}
	M \left( r_{\rm max} \right) = \sum_i m_i,
\end{equation}
where $m_i$ is the mass of the $i^{\rm th}$ dark matter particle.

We should make an initial guess at our parameters in order to find the maximum likelihood point.  We take the mass of the halo from the halo finder, $M_{\rm h}$, and estimate the concentration, $c_{\rm h} = r_{\rm vir} / r_s$, using the mass--concentration relation \cite{Child_2018}
\begin{equation}
\label{eq:mass_conc}
	c_{\rm h,  \, guess} = 57.6 \left(\frac{M_{\rm h}}{M_{\odot}} \right)^{-0.078}.
\end{equation}
We find that an initial guess of $r_{\rm s}$ equal to the 75${}^{\rm th}$ percentile of $r_{i}$ divided by $c_{\rm h, \, guess}$ is appropriate. Finally, we need to choose a value of $r_{\rm max}$ within which we fit our NFW profile. We make the simplest choice, setting $r_{\rm max}$ to be the virial radius, $r_{\rm vir}$, as output by the halo finder.

As an alternative to this method, we consider a parameterisation where the inner power law slope can vary. We consider a more general density profile
\begin{equation}
\label{eq:gennfw_profile}
	\rho_{\rm \Gamma} \left( r \right) = \frac{\rho_0}{\left( r / r_{\rm s} \right)^{-\Gamma} \left( 1 + r / r_{\rm s} \right)^{3+\Gamma}} ,
\end{equation}
which, like NFW, scales as $r^{-3}$ at large radii, but is allowed to have a different inner power law slope, $\Gamma$. Comparing to \autoref{eq:Power law density}, we see that for $r \ll r_{\rm s}$, we have $r_{\rm t} = r_{\rm s}$, $\rho_{\rm t} = \rho_0$ and $\beta = - \Gamma$.

We fit this using the same procedure as before, first fitting for $r_{\rm s}$ and $\Gamma$, and then finding $\rho_0$ by considering the total mass of particles. We now have the additional requirement that $\Gamma > -3$, so that $\lim_{r \to 0} M(r) = 0$.

\begin{figure*}
     \centering
    \subfloat[]{
         \centering
         \includegraphics[width=0.45\textwidth]{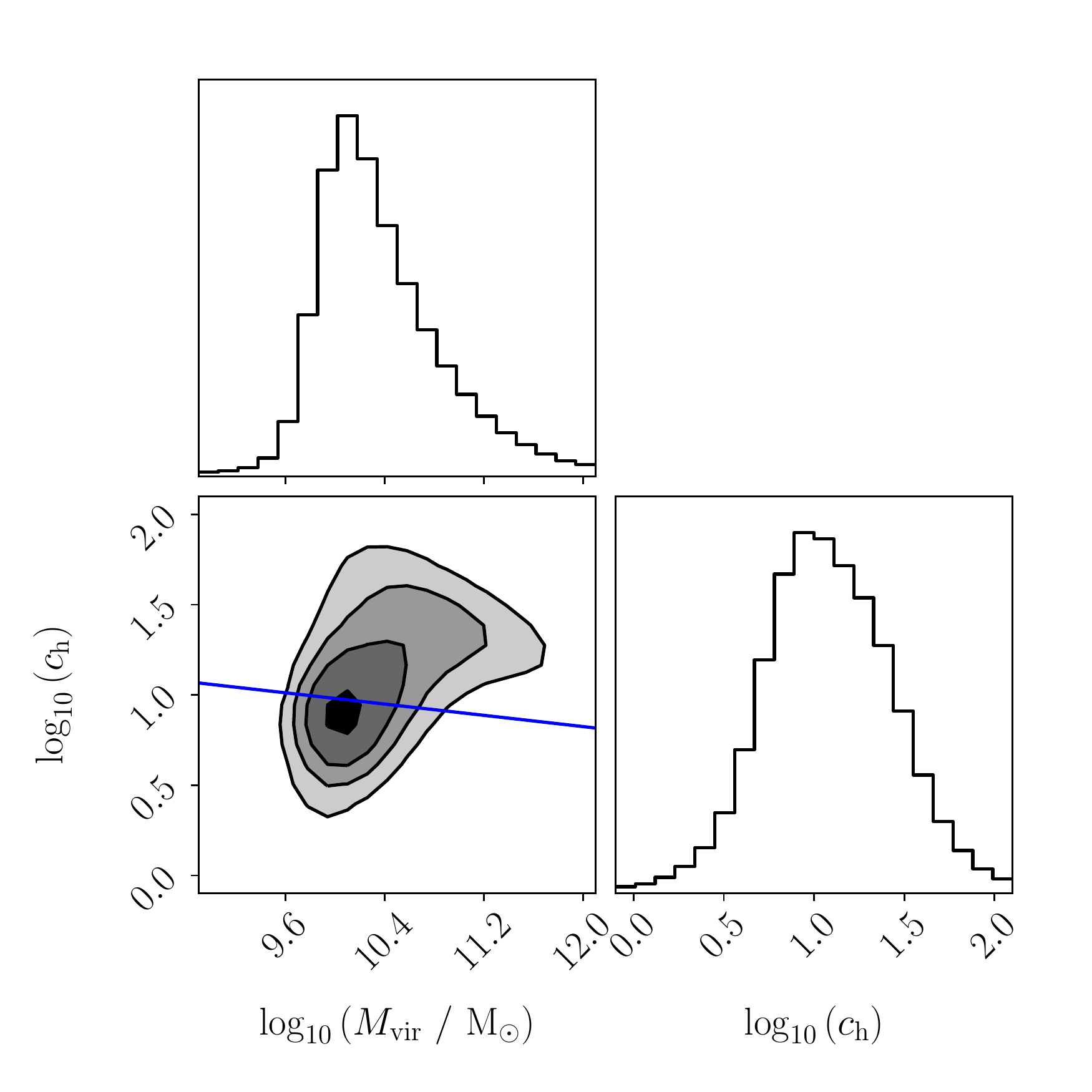}
         \label{subfig:nfw_parameters}
     }
     \subfloat[]{
         \centering
         \includegraphics[width=0.45\textwidth]{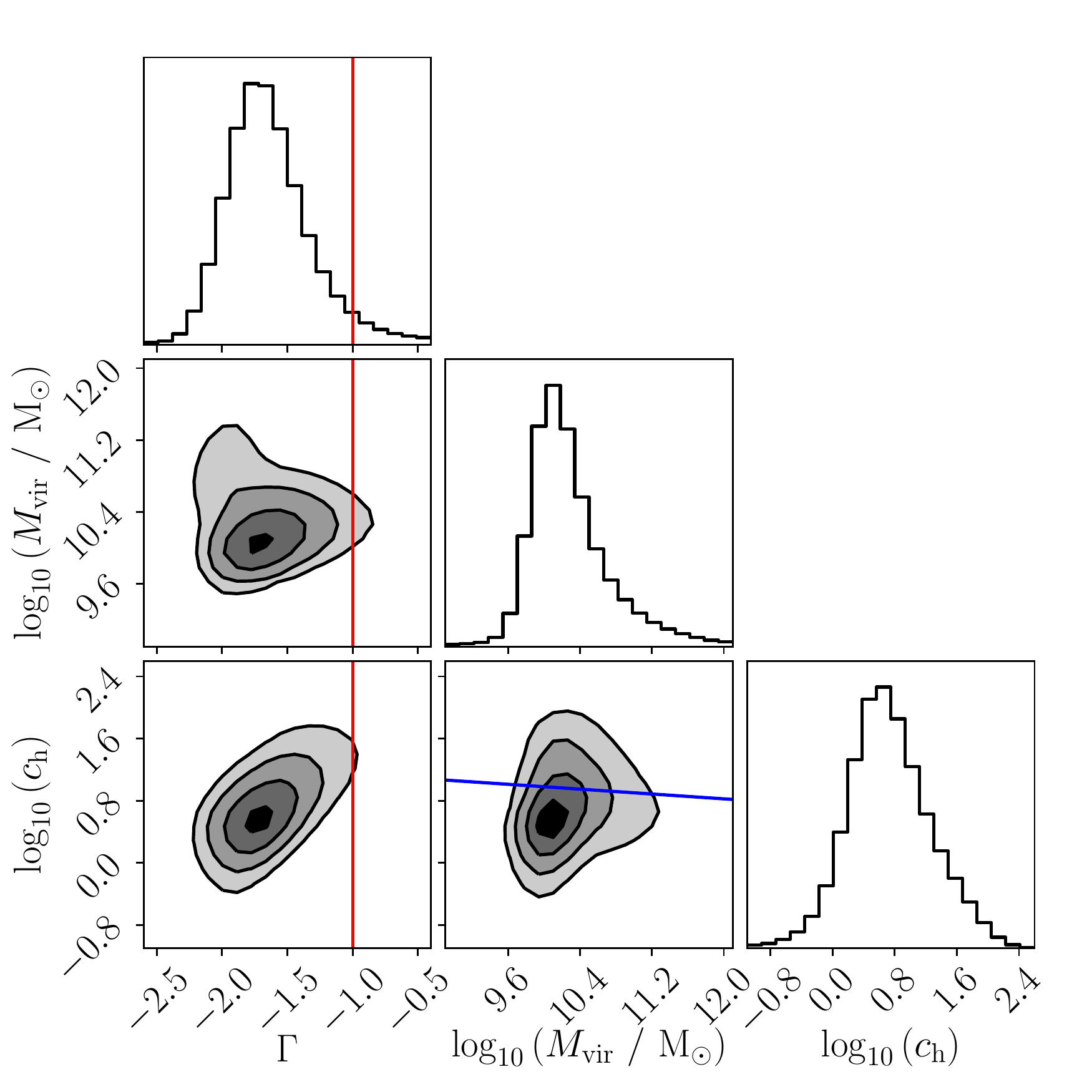}
         \label{subfig:gennfw_parameters}
     }
       \caption{Distributions of fitting parameters to halos in Horizon-AGN for \protect\subref{subfig:nfw_parameters} a NFW profile (\autoref{eq:nfw_profile}) and \protect\subref{subfig:gennfw_parameters} a NFW-like profile with a different inner power law slope (\autoref{eq:gennfw_profile}). Note that this is not a plot of posteriors, but a density map with one point per halo. The contours show the 1, 2 and $3\sigma$ levels of the distributions. The blue line is the mass-concentration relation given in \autoref{eq:mass_conc}. This falls within the $1\sigma$ region of \protect\subref{subfig:nfw_parameters}, whereas the more general profile favours a smaller concentration for a given virial mass. The red line in \protect\subref{subfig:gennfw_parameters} represents the NFW case ($\Gamma=-1$); most halos have steeper central density profiles than this ($\Gamma < -1$).
         }
		\label{fig:halo_profile_parameters}
\end{figure*}

The distributions of the fitting parameters for both of these profiles are shown in \autoref{fig:halo_profile_parameters}. We see that the majority of galaxies have $\Gamma < -1$, consistent with the conclusion of \cite{Peirani_2017} that the halos in Horizon-AGN have steeper density profiles near their centre than a NFW profile. We find that the mass-concentration relation of \autoref{eq:mass_conc} falls within the $1\sigma$ region of the distribution for the NFW profile, but the more general profile favours slightly lower concentrations.

\subsection{Modelling the offsets and warps}

\subsubsection{Gravitational and fifth force fields}

Horizon-AGN, by construction, does not include a fifth force. To check our inference method we confirm that it reconstructs no such force from the simulation, and calculate the fifth force constraints it would impose were it real data. To make the calculation computationally feasible, and to mimic the methods of \cite{f(R)_ruled_out,Desmond_2018_warp,Desmond_2018}, we consider two distinct contributions to the fifth force: i) a smoothed density field and ii) halos, which we assume are described by NFW profiles with the parameters obtained in \Cref{sec:Halo density profile}. To obtain the smoothed density field, we ignore all dark matter particles that are assigned to halos, and project the remaining dark matter particles and all of the star particles onto the same grid as the gas. This grid is defined to have $2^\ell$ cells per side across the full simulation volume. We assume that the density is Gaussian distributed in each cell,
\begin{equation}
	\rho_{\rm Gauss} \left( r \right) = \frac{M}{\left( 2 \pi \sigma^2 \right)^{\frac{3}{2}}} \exp \left( - \frac{r^2}{2 \sigma^2} \right),
\end{equation}
for
\begin{equation}
	\sigma = \frac{L_{\rm box}}{2^{\ell+1}},
\end{equation}
where $L_{\rm box}$ is the simulation box length.

Using these components, we first compute the Newtonian potential at the centre of each halo sourced by all mass within $\lambda_{\rm c}$ of that point. The contribution from a grid cell of mass $M$ of the smoothed density field at a distance $r$ is
\begin{equation}
\label{eq:Potential Gaussian}
    \Phi_{\rm ext, \, Gauss}\left( r \right) = - \frac{G_{\rm N}M}{r} \erf \left( \frac{r}{\sigma \sqrt{2}} \right).
\end{equation}
The contribution from a halo of virial mass $M_{\rm vir}$ and concentration $c_{\rm h}$ at a distance $r$ is
\begin{equation}
\label{eq:Potential NFW}
    \Phi_{\rm ext, \, NFW}\left( r \right) = - \frac{G_{\rm N}M_{\rm vir}}{r} \frac{\ln \left( 1 + c_{\rm h} r / r_{\rm vir} \right)}{\ln \left( 1 + c_{\rm h} \right) - c_{\rm h} / \left( 1 + c_{\rm h} \right)}.
\end{equation}
For each halo, we add a component due to self screening \cite{Cabre_2012,Zhao_2011b}
\begin{equation}
    \Phi_{\rm int} = - V_{\rm vir}^2,
\end{equation}
where $V_{\rm vir}$ is the virial velocity of the halo. The potential at a given halo is then
\begin{equation}
    \Phi =   \Phi_{\rm ext} + \Phi_{\rm int} = \Phi_{\rm ext, \, Gauss} + \Phi_{\rm ext, \, NFW} + \Phi_{\rm int},
\end{equation}
such that halos with $|\Phi|<\chi$ are unscreened, where $\chi$ is given by \autoref{eq:f(R) critical potential}.

To calculate the fifth force at the position of each halo, we sum the contributions from the unscreened halos and the unscreened regions of the smoothed density field within $4 \lambda_{\rm c}$ of the centre. To determine the Newtonian potential of the smoothed density field, we use a multidimensional piece-wise linear interpolator to interpolate the values of $\Phi_{\rm ext}$ from the halos to the grid points (as in \cite{f(R)_ruled_out}). Assuming there is no self screening of the smoothed density field, all grid points with $|\Phi_{\rm ext}| < \chi$ are then unscreened. By solving the time-independent massive Klein-Gordon equation (\autoref{eq:generalised Poisson} with $Z^{ij} = \delta^{ij}$), we find the magnitudes of the contributions from the Gaussian smoothed density field and NFW halos to be
\begin{widetext}
\begin{gather}
\begin{split}
\label{eq:a5 Gaussian}
	a_{\rm 5, \, Gauss} \left(r\right) &= - \frac{2 \alpha^2 G_{\rm N} M}{r^2} e^{-r/\lambda_{\rm c}} \left( 1 + \frac{r}{\lambda_{\rm c}} \right) e^{\frac{\sigma^2}{2\lambda_{\rm c}^2}} 
	\left[ \frac{1}{2} \left( 1 + \erf \left( \frac{r - \sigma^2/\lambda_{\rm c}}{\sigma \sqrt{2}} \right) \right) - \right. \\
	& \left. \frac{1}{2} \frac{1-r/\lambda_{\rm c}}{1+r/\lambda_{\rm c}} e^{2 r/\lambda_{\rm c}} \left( 1 - \erf \left( \frac{r + \sigma^2/\lambda_{\rm c}}{\sigma \sqrt{2}} \right) \right)  - \frac{2 r}{1+r/\lambda_{\rm c}} \frac{1}{\sqrt{2\pi}\sigma} \exp \left( - \frac{\left( r - \sigma^2/\lambda_{\rm c} \right)^2}{2\sigma^2} \right) \right],
\end{split} \\
\begin{split}
\label{eq:a5 NFW}
	& a_{\rm 5, \, NFW} \left(r\right) = - \frac{2 \alpha^2 G_{\rm N} M_{\rm vir}}{r^2} \frac{c_{\rm h} \left(1 + c_{\rm h} \right) e^{- \left(r + r_{\rm vir}/c_{\rm h} \right) / \lambda_{\rm c}}}{2 \left( r_{\rm vir} + c_{\rm h} r \right) \left( \left( 1 + c_{\rm h} \right) \ln \left( 1 + c_{\rm h} \right) - c_{\rm h} \right)} \left( e^{2 \left(r + r_{\rm vir}/c_{\rm h} \right) / \lambda_{\rm c}} \left( \frac{r}{\lambda_{\rm c}} - 1 \right) \left( \frac{r_{\rm vir}}{c_{\rm h}} + r \right)  \Ei \left( - \frac{r + r_{\rm vir} / c_{\rm h}}{\lambda_{\rm c}} \right) - \right. \\
	& \left.\left( \frac{r}{\lambda_{\rm c}} + 1 \right) \left( \frac{r_{\rm vir}}{c_{\rm h}} + r \right)  \Ei \left( \frac{r + r_{\rm vir} / c_{\rm h}}{\lambda_{\rm c}} \right) + 2 e^{r_{\rm vir} / \left( \lambda_{\rm c} c_{\rm h} \right)} \left( e^{r / \lambda_{\rm c}} r +\left( \frac{r}{\lambda_{\rm c}} + 1 \right) \left( \frac{r_{\rm vir}}{c_{\rm h}} + r \right) \left( \gamma + \ln \left( \frac{r_{\rm vir}}{\lambda_{\rm c} c_{\rm h}} \right) \right) \right) \right),
\end{split}
\end{gather}
\end{widetext}
where $\Ei$ is the exponential integral function and $\gamma$ is Euler’s gamma constant.

We choose $\ell = 7$, which corresponds to a spatial resolution of $\Delta x \sim 1.1 {\rm \, Mpc}$, comparable to the $1.4 {\rm \, Mpc}$ resolution used for the smoothed density field in \cite{f(R)_ruled_out}. As discussed in \cite{Desmond_2018}, if $\lambda_{\rm c}$ is less than a few $\Delta x$, the discretisation of the smoothed density field can lead to excessive shot noise. Hence, in the cases where $\lambda_{\rm c} < R_{\rm thresh} $, we evaluate the potential and acceleration from the smoothed density field at a cutoff of $R_{\rm thresh}$ and $4 R_{\rm thresh}$ respectively, then correct our results as
\begin{equation}
    \Phi_{\rm ext, \, Gauss} \left( \lambda_{\rm c} \right) = \left( \frac{\lambda_{\rm c}}{R_{\rm thresh}} \right)^2 \Phi_{\rm ext, \, Gauss} \left( R_{\rm thresh} \right),
\end{equation}
and 
\begin{equation}
    \bm{a}_{\rm Gauss} \left( \lambda_{\rm c} \right) = \left( \frac{\lambda_{\rm c}}{R_{\rm thresh}} \right)^2 \bm{a}_{\rm Gauss} \left( R_{\rm thresh} \right),
\end{equation}
where, as in \cite{f(R)_ruled_out}, we choose $R_{\rm thresh} = 3.5 {\rm \, Mpc}$.

\subsubsection{Modelling the halo restoring force}
\label{sec:Modelling the halo restoring force}

We focus on the power-law region of \autoref{eq:Power law density} ($r < r_{\rm t}$), where the mass enclosed within a radius $r$ is given by
\begin{equation}
    M\left(r\right) = \frac{4 \pi \rho_{\rm t}}{3 - \beta} r_{\rm t}^{\beta} r^{3 - \beta}.
\end{equation}
Using \cref{eq:General offset}, we then find the predicted offset to be
\begin{equation}
    \bm{r}_{\star} = \left( a_5 \frac{\Delta G}{G_{\rm N}^2} \frac{3 - \beta}{4 \pi \rho_{\rm t}} r_{\rm t}^{-\beta} \right)^{\frac{1}{1-\beta}} \hat{a}_5,
\end{equation}
and, using \cref{eq:General z,eq:General w1}, the predicted warp parameter is
\begin{equation}
\label{eq:power law w1}
    w_1 = \frac{- \beta \left( 3 - \beta \right)}{\left( 1 + \beta \right) \left( 2 + \beta \right)} \frac{\Delta G}{G_{\rm N}^2} \frac{1}{4 \pi \rho_{\rm t}} \frac{\left( 3 R_{\rm eff} / r_{\rm t} \right)^{\beta}}{3 R_{\rm eff}} \bm{a}_5 \cdot \hat{z}.
\end{equation}
One may be concerned that a larger $\Delta G / G_{\rm N}$ would result in a smaller offset for $\beta > 1$. By considering a small perturbation about equilibrium in this case, one finds that the offset is unstable and thus the predicted signal is either zero or infinite. We therefore set $\bm{r}_{\star} = \bm{0}$ whenever $\beta \geq 1$. We evaluate these predicted signals for $\Delta G / G_{\rm N} = 1$ to create a ``template'' signal, which can then be multiplied by appropriate functions of $\Delta G / G_{\rm N}$ to obtain the corresponding prediction (see \autoref{eq:signal likelihood} below).

\subsection{Selection criteria}

In this section we summarise the selection criteria used to obtain the samples for our inference, having justified these in the preceding sections. The fiducial values for these cuts are also shown in \autoref{tab:parameter_summary}.

\begin{table*}
    \caption{The fiducial parameters used in the offset and warp inferences, as described in the text. The first six parameters are used in both inferences, whereas the next two are only used for warps, and the final four are for the offset analysis.}
    \label{tab:parameter_summary}
    \centering
    \begin{tabular}{l|l|l}
    \textbf{Parameter} & \textbf{Description} & \textbf{Value} \\
    \hline
    $\lambda_{\rm c}$ & Compton wavelength of fifth force field. &  $\lambda_{\rm c} \in [0.4, 7.6] {\rm \, Mpc}$\\
    Halo profile & Type of density profile used in fit. &  NFW  \\
    $M_{\rm g, \, crit}$ & Minimum galaxy mass. & $2 \times 10^{9} {\, \rm M_{\sun}}$ \\
    $M_{\rm h, \, crit}$ & Minimum halo mass. & $8 \times 10^{10} {\, \rm M_{\sun}}$ \\
    $r_{\rm t}$ & Where to start core. & $r_{\rm s}$   \\
    $\beta$ & Power law slope of core. & 0.5   \\
    \hline$
    \epsilon_{\rm crit} $ & Minimum ellipticity of S\'{e}rsic fit of galaxy. & 0.5 \\
    $N_{\rm eff}$ & Within how many $R_{\rm eff}$ to calculate warp. & 3   \\
    \hline
    $N_{\star}$ & Within how many $R_{\rm eff}$ to calculate gas centre of mass. &    4 \\
    $R^2_{\rm crit}$ & Minimum value of $R^2$ for Gaussian fit to gas density to say fit has converged. & 0.6 \\
    $N_{\rm g}$ & Maximum number of gas length scales ($l_{\rm g}$) for which gas is associated to a galaxy. & 4 \\
    $N_{\rm nn}$ & Closest a nearest neighbouring galaxy can be as a multiple of the sum of their $R_{\rm eff}$. & 4 \\
    \end{tabular}
\end{table*}

\subsubsection{Warp sample}

\begin{figure}
	 \includegraphics[width=\columnwidth]{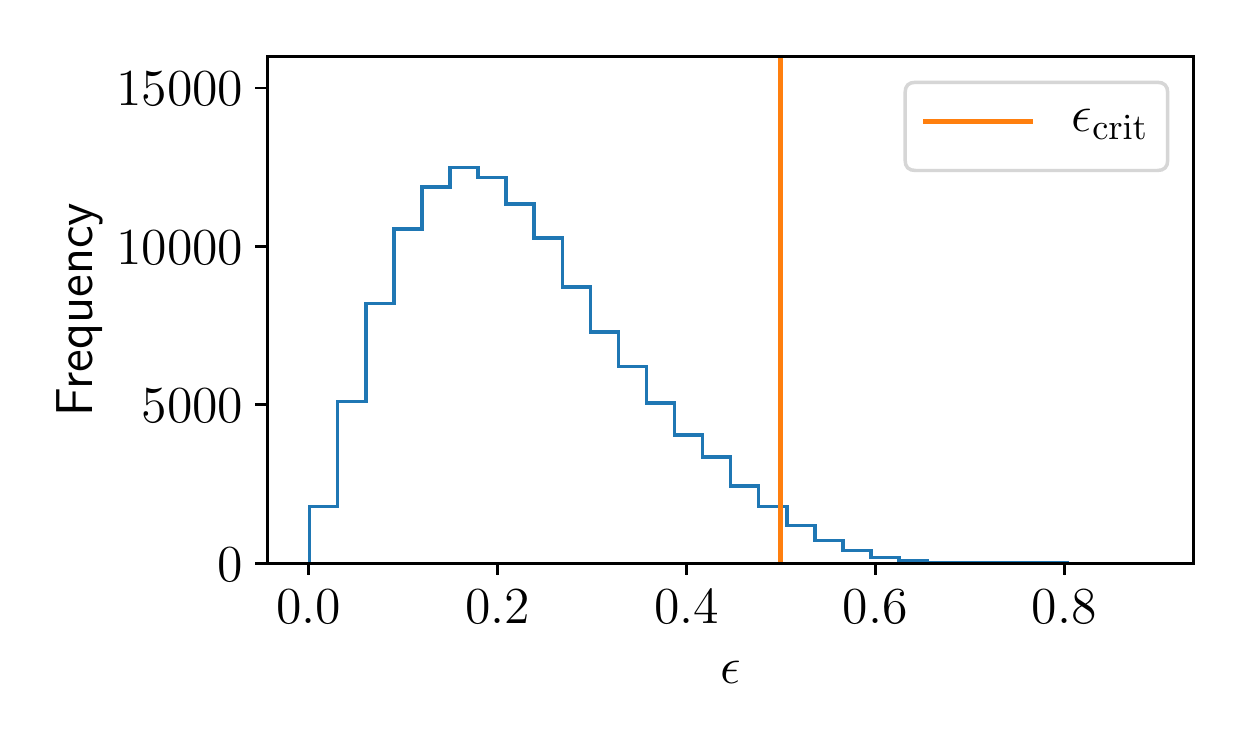}
	 \caption{\label{fig:ellipticity_distribution} The distribution in ellipticity, $\epsilon$, of the galaxies in Horizon-AGN. The vertical line shows the cut at $\epsilon_{\rm crit}=0.5$ used in \cite{Desmond_2018_warp}, where only galaxies to the right of the line are used. We use the same cut here, reducing our Horizon-AGN sample to 2,990 galaxies.}
\end{figure}

As noted in \Cref{sec:Stellar warp}, the S\'{e}rsic fit converges for over 99 per cent of our galaxies, reducing the sample from 126,361 to 125,346. In \autoref{fig:ellipticity_distribution} we plot the distribution of ellipticity, $\epsilon$, for these galaxies. The vertical line at $\epsilon=\epsilon_{\rm crit}=0.5$ is the cut used in \cite{Desmond_2018_warp} to keep only disk-like galaxies, where we keep those with $\epsilon>0.5$. Since the finite spatial resolution will dilate disks with scale heights below $1 {\rm \, kpc}$ \cite{Welker_2017} and thus decreases their ellipticity, this criterion dramatically reduces our sample to 2,990. A further 273 galaxies are removed for having stellar masses below $2 \times 10^{9} {\rm \, M_{\sun}}$ and 47 more do not have finite warp values due to having columns of zero intensity in their mock images. 22 of these galaxies do not have associated halos. 54 of the remaining galaxies have a halo mass below $8 \times 10^{10} {\rm \, M_{\sun}}$, which when rejected leaves a final sample of 2,594 galaxies. This should be compared to the 4,139 galaxies used in \cite{f(R)_ruled_out}.

\subsubsection{Offset sample}

Starting with the 117,099 galaxy+halo pairs, we discard the 14,041 galaxies where the Gaussian fit to the surrounding gas does not converge, as detailed in \Cref{sec:Gas-star offset}. The gas--star offset is greater than $N_{\rm g} l_{\rm g}$ for 430 galaxies, where $N_{\rm g} = 4$, so these are also eliminated from the sample. 1,923 galaxies have nearest neighbours which are too close to make the gas--star offset measurement reliable and 44,514 have masses below $2 \times 10^{9} {\rm \, M_{\sun}}$, reducing our sample to 56,132. Imposing a minimum halo mass of $8 \times 10^{10} {\rm \, M_{\sun}}$ leaves a final sample of 38,042 galaxies, to be compared to 15,634 in \cite{f(R)_ruled_out}.

\subsection{Likelihood model}
\label{sec:Likelihood model}

Now for each galaxy, $i$, in the offset and warp samples, we have both an ``observed'' (simulated), $y_{i}$, and template, $\bar{y}_{i}$, signal, where $y \in \{ r_{\star, \rm \alpha}, r_{\star, \rm \delta}, w_1 \}$. The likelihood for the observed signal is then
\begin{equation}
\label{eq:signal likelihood}
    \begin{split}
    \mathcal{L}_{i} & \left( y_i | \Delta G / G_{\rm N}, \lambda_{\rm c}, \sigma_{y, i} \right) \\
    &  = \frac{1}{\sqrt{2 \pi} \sigma_{y, i}} \exp \left( - \frac{ \left(y_{i} - \bar{y}_{i} \left(\Delta G / G_{\rm N} \right)^{\tau} \right)^2}{2\sigma_{y, i}^2} \right),
    \end{split}
\end{equation}
where $\tau = 1$ for the warp inference, and $\tau = (1 - \beta)^{-1}$ for the offset inference. The noise parameters, $\bm{\Omega}_y = \{ \sigma_{y, i} \}$, characterise the uncertainty on each of the parameters, and we use the same $\sigma_{y,i}$ for $y=r_{\star, \rm \alpha}$ and $y=r_{\star, \rm \delta}$. A key assumption of \cite{f(R)_ruled_out} is that these parameters are either constant for all galaxies or only depend linearly on the distance between the observer and galaxy, $d_{i}$. For our fiducial analysis we choose a constant $\bm{\Omega}_y$ for all galaxies, as the spatial resolution of Horizon-AGN means that a physical, as opposed to angular, uncertainty is appropriate for the offset inference, and we find no systematic trend between $w_1$ and distance. In \Cref{sec:Validity of halo density model} we investigate the validity of these assumptions.

Assuming the galaxies are independent, we find the likelihood of the set of observed $y$ to be
\begin{equation}
    \mathcal{L} \left( y | \Delta G / G_{\rm N}, \lambda_{\rm c}, \bm{\Omega}_{y} \right) = \prod_{i} \mathcal{L}_{i} \left( y_i | \Delta G / G_{\rm N}, \lambda_{\rm c}, \sigma_{y,i} \right).
\end{equation}
We also treat the signals as independent, such that the total likelihood of our dataset $\mathcal{D}$ is
\begin{equation}
    \mathcal{L} \left( \mathcal{D} | \Delta G / G_{\rm N}, \lambda_{\rm c}, \bm{\Omega} \right) = \prod_{y} \mathcal{L} \left( y | \Delta G / G_{\rm N}, \lambda_{\rm c}, \bm{\Omega}_{y} \right), 
\end{equation}
for $\bm{\Omega} = \{\bm{\Omega}_y\}$. We consider both the warp and offset samples separately and combined, so that for the above product we have three choices: $y = w_1$ (warp inference), $y \in \{ r_{\star, \alpha}, r_{\star, \delta} \}$ (offset inference), and $y \in \{ w_1, r_{\star, \alpha}, r_{\star, \delta} \}$ (combined inference). Finally, given some prior on $\Delta G$, $\lambda_{\rm c}$ and $\bm{\Omega}$, $P\left(\Delta G, \lambda_{\rm c}, \bm{\Omega} \right)$, we use Bayes' theorem to obtain
\begin{equation}
	P\left(\Delta G, \lambda_{\rm c}, \bm{\Omega}| \mathcal{D} \right) = \frac{\mathcal{L}\left(\mathcal{D}|\Delta G, \lambda_{\rm c}, \bm{\Omega} \right) P\left(\Delta G, \lambda_{\rm c}, \bm{\Omega} \right)}{P\left(\mathcal{D}\right)},
\end{equation}	
where $P(\mathcal{D})$ is the constant probability of the data for any $\{ \Delta G, \lambda_{\rm c}, \bm{\Omega} \}$. 

Imposing the priors $\Delta G / G_{\rm N} \geq 0$ and $\sigma_{y,i} > 0 \, \forall \, y, i$, and using the \textsc{emcee} sampler \cite{emcee}, we now derive posteriors on $\Delta G / G_{\rm N}$ and the noise model parameters at fixed $\lambda_{\rm c}$. We sample with 32 walkers and terminate the chain when the estimate of the autocorrelation length changes by less than 1 per cent per iteration and the chain is at least 50 autocorrelation lengths long.

\section{Simulated constraints}
\label{sec:Constraints}

\begin{figure}
	 \includegraphics[width=\columnwidth]{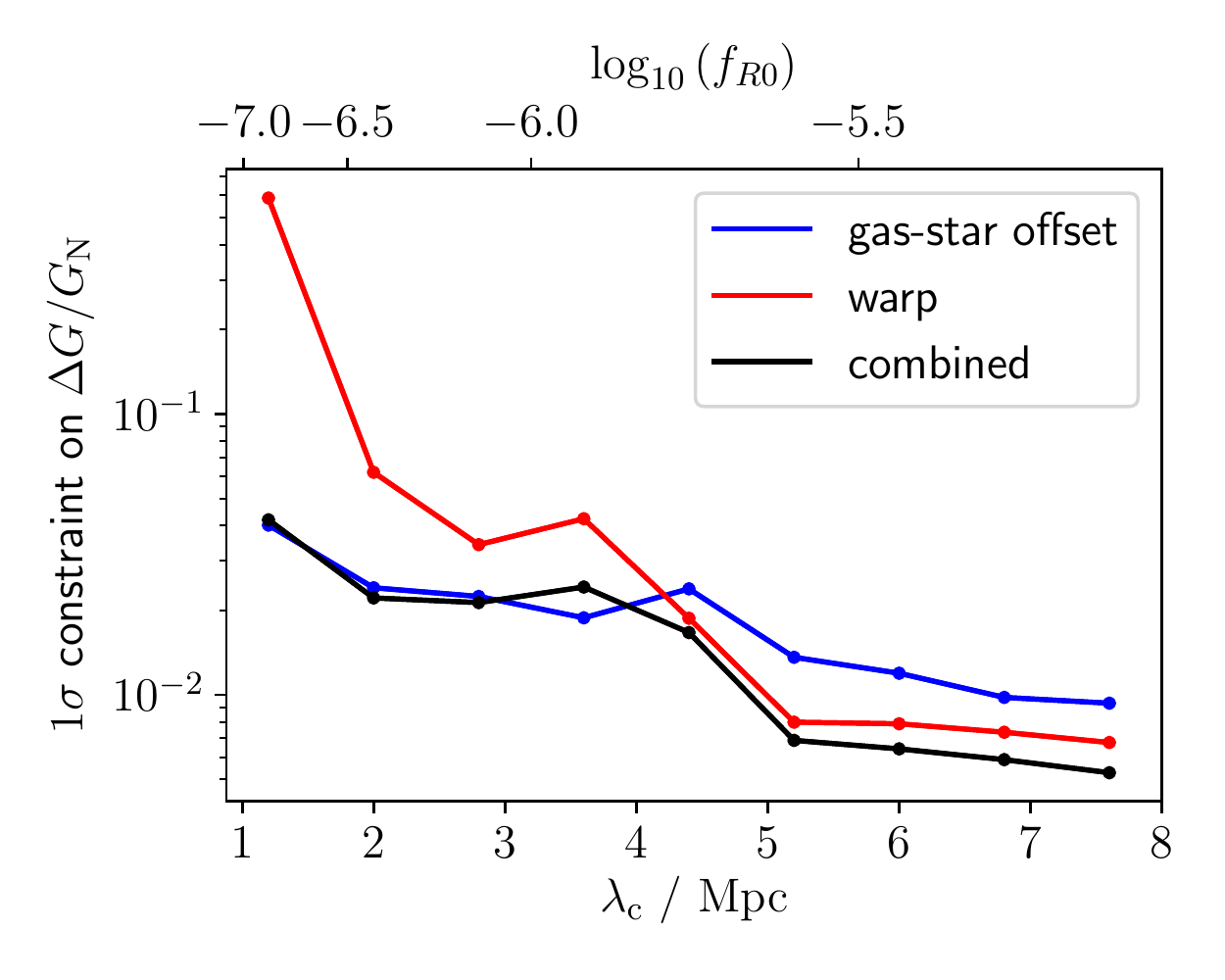}
	 \caption{\label{fig:hagn_constraints}$1 \sigma$ constraints on $\Delta G / G_{\rm N}$ as a function of $\lambda_{\rm c}$ or $n=1$ Hu-Sawicki $f_{R0}$ for the gas--star offset (blue) and warp (red) analyses, and their combination (black), using simulated galaxies from Horizon-AGN.
	 }
\end{figure}

In \autoref{fig:hagn_constraints} we plot the $1 \sigma$ constraints on $\Delta G / G_{\rm N}$ as a function of $\lambda_{\rm c}$ for the warp and offset samples separately, as well as the joint constraint obtained from multiplying the likelihoods. We find the same qualitative results as \cite{f(R)_ruled_out}; $\Delta G / G_{\rm N}$ is consistent with zero, and the strength of the constraint improves with increasing $\lambda_{\rm c}$. The warp inference is weaker than the gas--star offset inference at small $\lambda_{\rm c}$ because there are far fewer galaxies in the warp sample; repeating the gas--star offset inference with the same number of galaxies as the warp sample results in comparable constraints for both signals. We find that all halos in our sample are screened for $\lambda_{\rm c} = 0.4 {\rm \, Mpc}$, so we cannot achieve a constraint for this or lower $\lambda_{\rm c}$.

\begin{figure}
	 \includegraphics[width=\columnwidth]{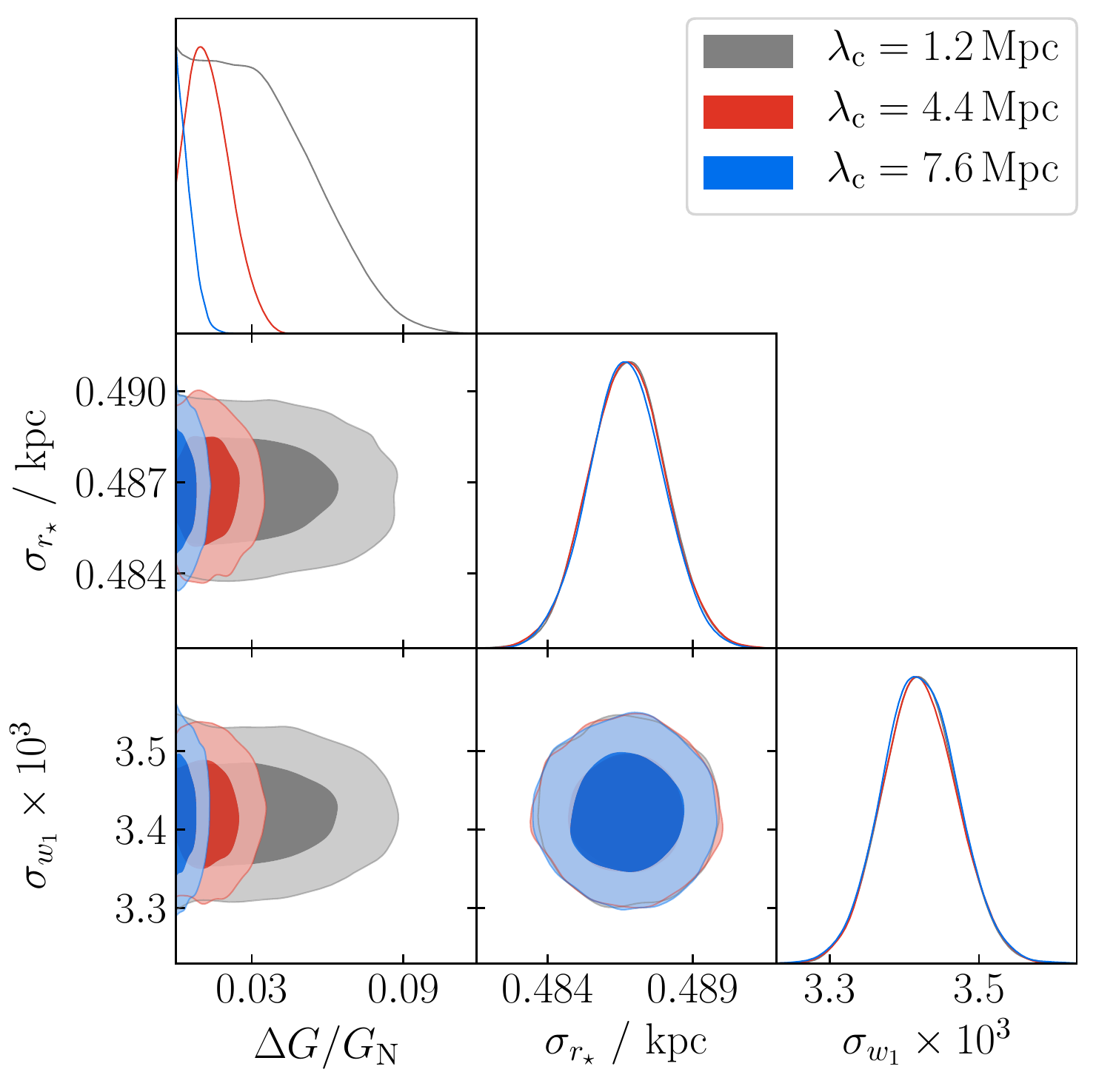}
	 \caption{\label{fig:hagn_corner} Constraints on $\Delta G / G_{\rm N}$ at different $\lambda_{\rm c}$, from the combined inference of the gas--star offsets and galaxy warps in galaxies from Horizon-AGN, along with the noise parameters $\sigma_{r_{\star}}$ and $\sigma_{w_1}$.}
\end{figure}

For $\lambda_{\rm c} = 1.2, \, 4.4$ and $7.6 {\rm \, Mpc}$, we plot the posterior distributions from the combined inference in \autoref{fig:hagn_corner}. We see that the typical scale of the offsets is $\sigma_{r_{\star}} \sim 0.5 {\rm \, kpc}$, which is approximately half the spatial resolution of Horizon-AGN. We also see that $w_1$ in Horizon-AGN is approximately five times larger than found in \cite{f(R)_ruled_out}. As previously noted, the finite spatial resolution inflates disks with scale heights below $1 {\rm \, kpc}$ \cite{Welker_2017}. This leads to larger absolute fluctuations in the luminosity-weighted $z$ position of the disk, justifying the increased $w_1$. It is therefore reasonable to suppose that the magnitudes of both of these noise parameters are set by the resolution of the simulation, so that these are upper limits for the true theoretical predictions. However, it is not the magnitude of the signals which we wish to determine here, but their correlations with other galaxy properties, which one would expect not to change significantly with improved resolution. Since the absolute level of noise will have an impact on the constraints in \autoref{fig:hagn_corner}, we do not compare these directly to \cite{f(R)_ruled_out}, but rather consider the relative change to our constraints if we alter the noise model.

\section{Validity of noise model}
\label{sec:Validity of noise model}

We now are ready to address the first two questions in \Cref{sec:General method}; are there unaccounted-for correlations in the noise, and, if so, do these impact the constraints?
As noted in \Cref{sec:Likelihood model}, both our fiducial analysis and \cite{Desmond_2018_warp,f(R)_ruled_out} assumed that the noise in the warp inference is a constant for all galaxies, whereas for the offset inference we assumed a constant spatial uncertainty and \cite{Desmond_2018,f(R)_ruled_out} assumed a constant angular uncertainty. 
Assessing whether the observable is correlated with parameters used to construct the template signal due to baryonic physics (i.e. in the simulation) is analogous to asking whether the former can be predicted from the latter through some function. If not, an empirical noise model in which such correlations are absent is sufficient. Otherwise a more sophisticated noise model may be required.

To determine the halo parameters to use, we fit each simulated halo with a NFW and the NFW-like profile with the inner power law slope as a free parameter, and choose whichever fit minimises the Bayesian Information Criterion,
\begin{equation}
\label{eq:BIC}
	{\rm BIC} = \mathcal{K} \log \mathcal{N} - 2 \log \hat{\mathcal{L}},
\end{equation}
for $\mathcal{K}$ model parameters, $\mathcal{N}$ halo particles, and maximum likelihood value $\hat{\mathcal{L}}$. We then have the characteristic density, $\rho_0$, scale radius, $r_{\rm s}$, inner power law slope, $\Gamma$, and virial radius, $r_{\rm vir}$, for each halo. We combine these with the distance of the galaxy from the centre of the box, $d_{i}$, screening potential, $\Phi$, and magnitude of the fifth force field at a given $\lambda_{\rm c}$, $|\bm{a}_5|$, to obtain the set of parameters that we will consider correlations of the noise with.

\begin{figure*}
     \centering
     \subfloat[]{
         \centering
         \includegraphics[width=0.45\textwidth]{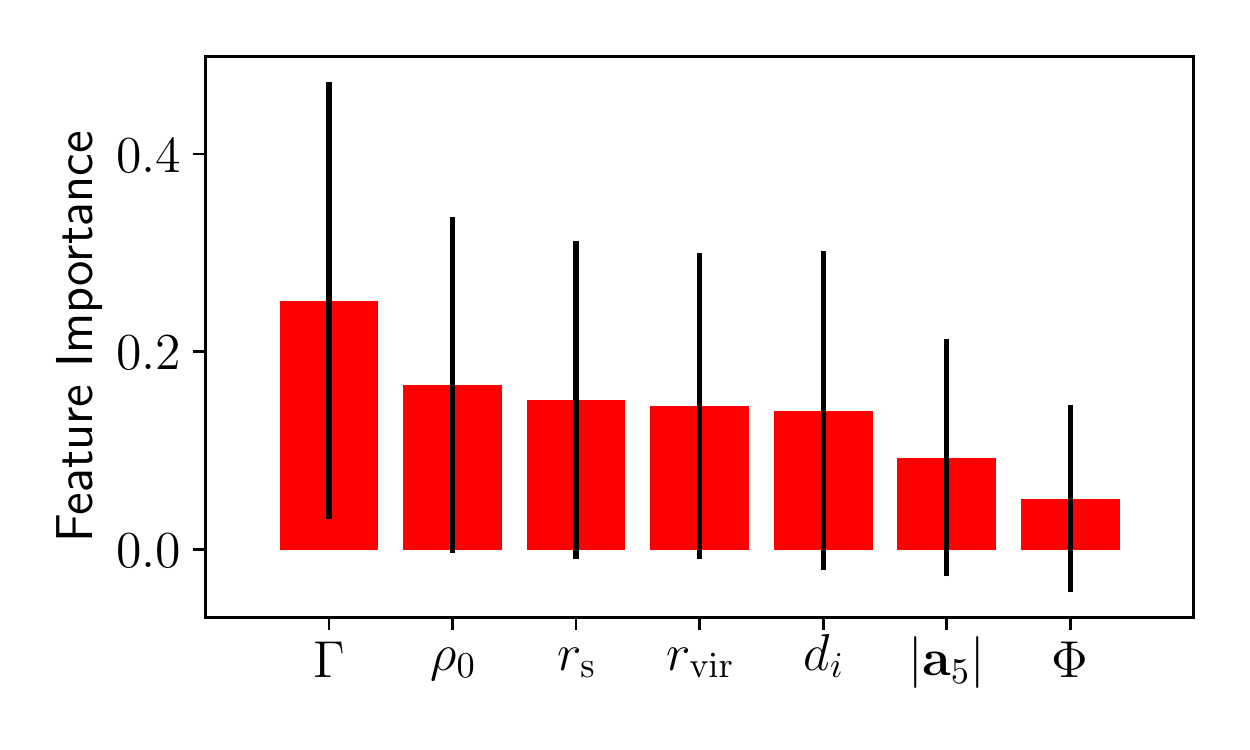}
         \label{subfig:feature_importance_warps}
     }
     \subfloat[]{
         \centering
         \includegraphics[width=0.45\textwidth]{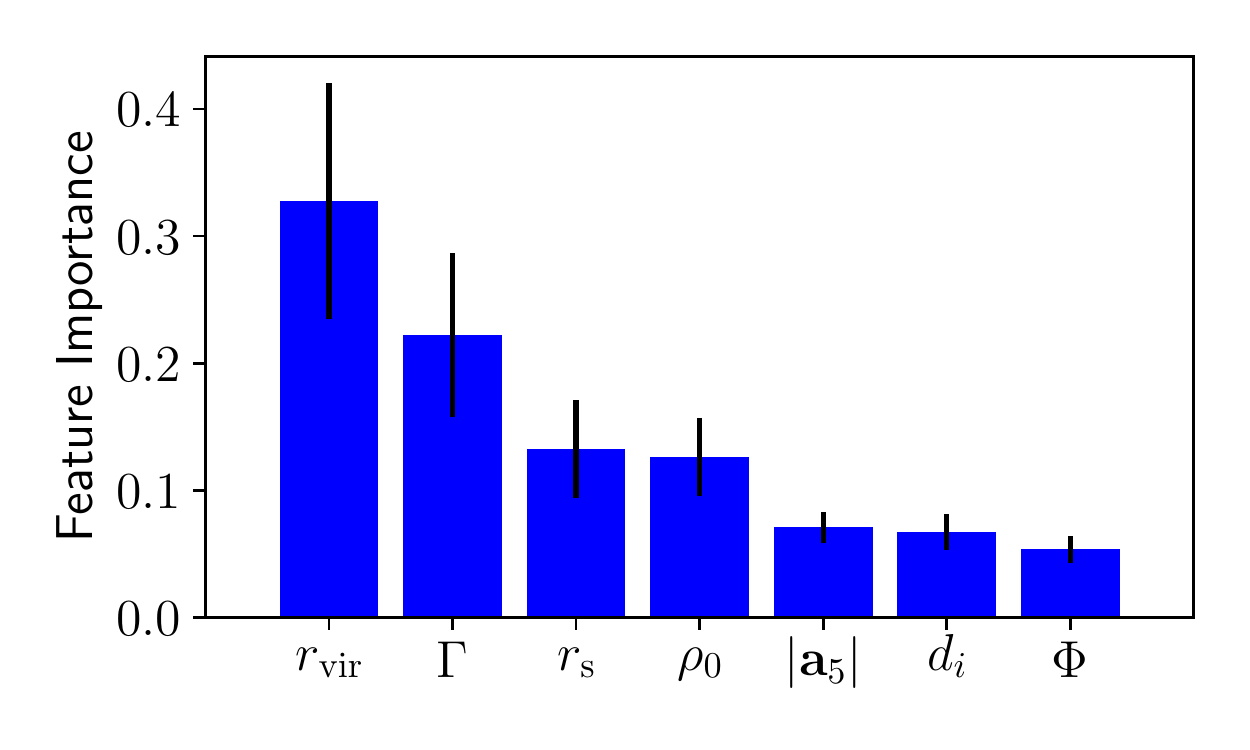}
         \label{subfig:feature_importance_offsets}
     }
        \caption{Feature importances for predicting \protect\subref{subfig:feature_importance_warps} warps, $w_1$, and \protect\subref{subfig:feature_importance_offsets} gas--star offsets, $|\bm{r_{\star}}|$, in the simulation from the variables relevant to the fifth force prediction using optimised Random Forest regressors. The bars are the impurity-based feature importances (normalised so that their sum is one), and the lines give the inter-tree variability. $w_1$ is found to be uncorrelated with any such variable as each feature importance is consistent with zero. Conversely, the gas--star offset is correlated with several properties of the host halos.}
		\label{fig:feature_importances}
\end{figure*}

To determine the level of correlation, we calculate the feature importances from optimised Random Forests for the prediction of $w_1$ and $\left|\bm{r}_\star\right|$ from these parameters. The feature importance gives the total decrease in node impurity due to that feature, normalised so that the sum of feature importances is one. The most important features---those that correlate most strongly with the signal---have the largest feature importances, and the inter-tree variability (shown by the black lines in \autoref{fig:feature_importances}) indicates the level of uncertainty on these values.

\begin{figure*}
    \centering
    \includegraphics[width=\textwidth]{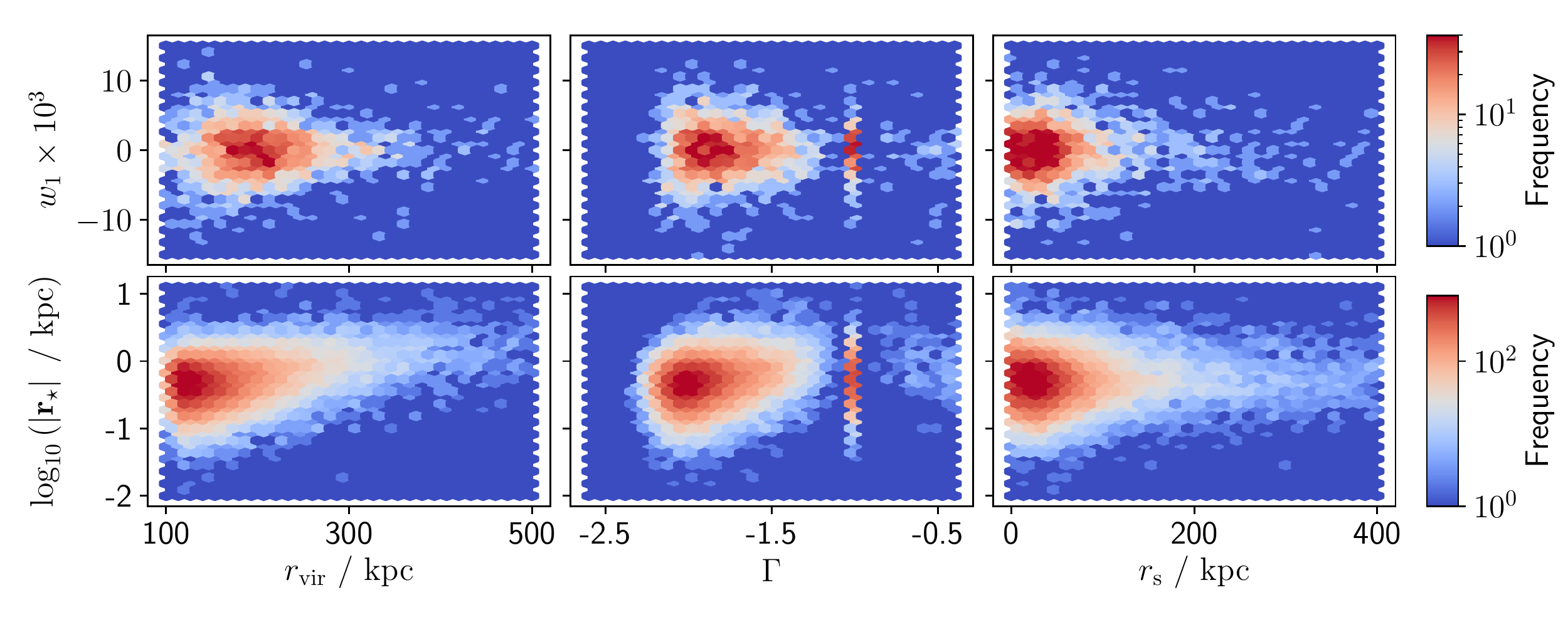}
    \caption{Two-dimensional histograms of the simulated signals and the most important features, as given in \autoref{fig:feature_importances}, for determining the gas--star offset in Horizon-AGN. We see that these features have little correlation with the warp statistic, but there is a clear correlation between $r_{\rm vir}$ and the magnitude of the gas--star offset. The high-intensity band at $\Gamma=-1$ contains the galaxies whose halos prefer a NFW fit over the more general profile according to the Bayesian Information Criterion.
    }
	\label{fig:hexbin_corr}
\end{figure*}

\subsection{Correlation of warps with galaxy and halo properties}

In \autoref{subfig:feature_importance_warps} we plot the feature importances for the prediction of $w_1$ from the parameters listed above that are used to make the template signal. We see that all features are equally (un)important and find that the regressor is unable to predict $w_1$ reliably, with a cross-validated score of 0.02. This is also evident in the two-dimensional histograms plotted in \autoref{fig:hexbin_corr}, where we see little correlation between $w_1$ and the parameters considered. It is therefore appropriate to assume uncorrelated noise in the inference, justifying the model of \cite{Desmond_2018_warp,f(R)_ruled_out}.

\subsection{Correlation of offsets with galaxy and halo properties}

The case of gas--star offsets is more interesting, as \autoref{subfig:feature_importance_offsets} indicates that the properties of the halo are important in predicting the measured value, and indeed we obtain a higher cross-validated score of 0.45. We find that the two most important features are $r_{\rm vir}$ and $\Gamma$, and the correlations of the signals with these parameters are clearly visible in the two-dimensional histograms of \autoref{fig:hexbin_corr}. The relationship between $r_{\rm vir}$ and $\left|\bm{r}_{\star}\right|$ is not linear, indicating that the offset does not solely arise via scale-invariant processes.

To determine whether any correlation with these parameters affects our constraint, we now allow $\sigma_{r_\star}$ to vary with one of these parameters, which we denote $p$. We sort our galaxies into bins of increasing $p$ such that each bin contains the same number of members, except in the case $p = \Gamma$, where we have one bin which is larger, containing all galaxies that are best-fit by NFW profiles ($\Gamma=-1$). We repeat our inference with a universal $\Delta G / G_{\rm N}$, but with a different $\sigma_{r_\star}$ for each bin, and find that the fitted $\sigma_{r_\star}$ are independent of both $\Delta G / G_{\rm N}$ and $\lambda_{\rm c}$.

\begin{figure*}
     \centering
     \subfloat[]{
         \centering
         \includegraphics[width=0.45\textwidth]{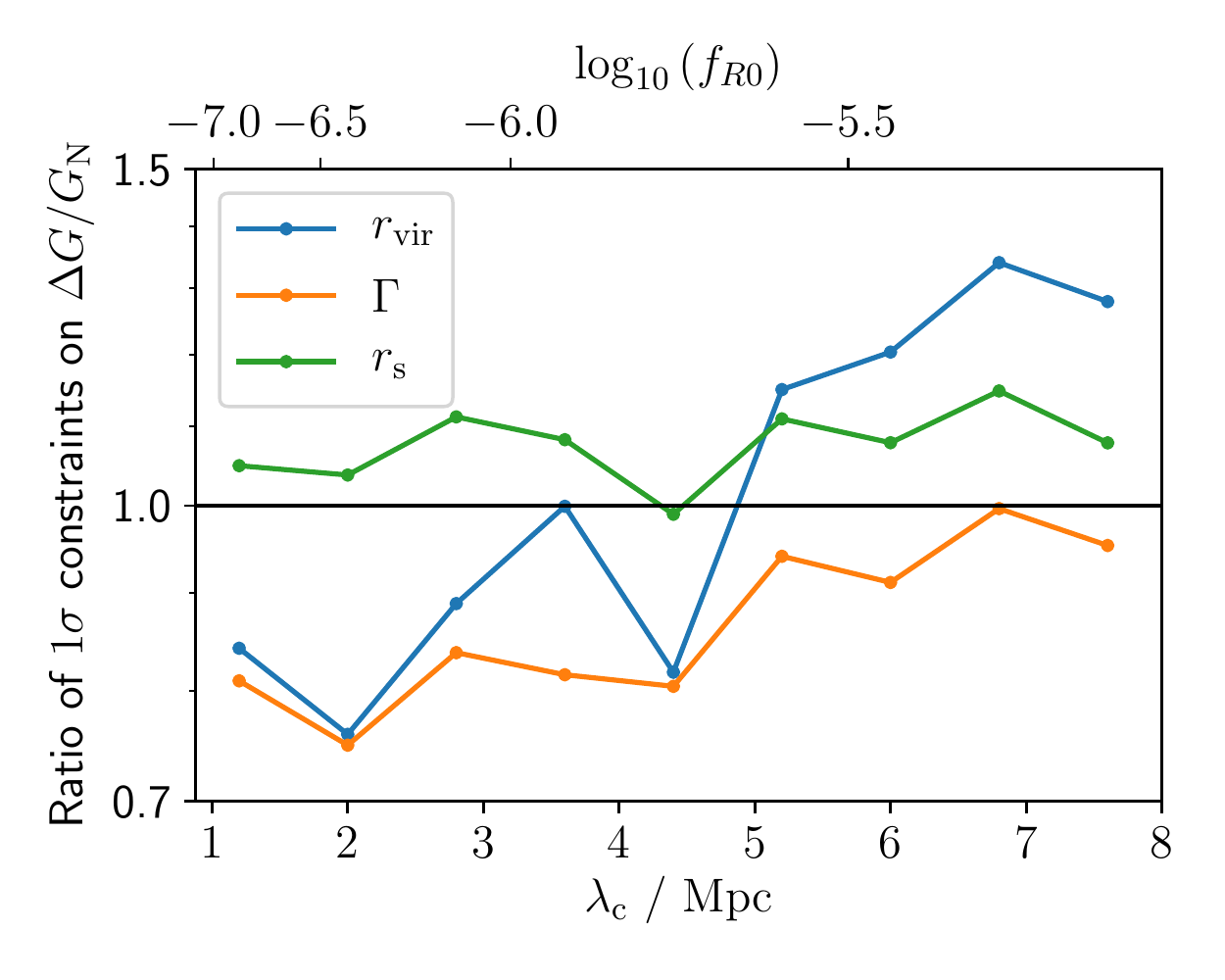}
         \label{subfig:hagn_constraints_binned}
        }
     \subfloat[]{
         \centering
         \includegraphics[width=0.45\textwidth]{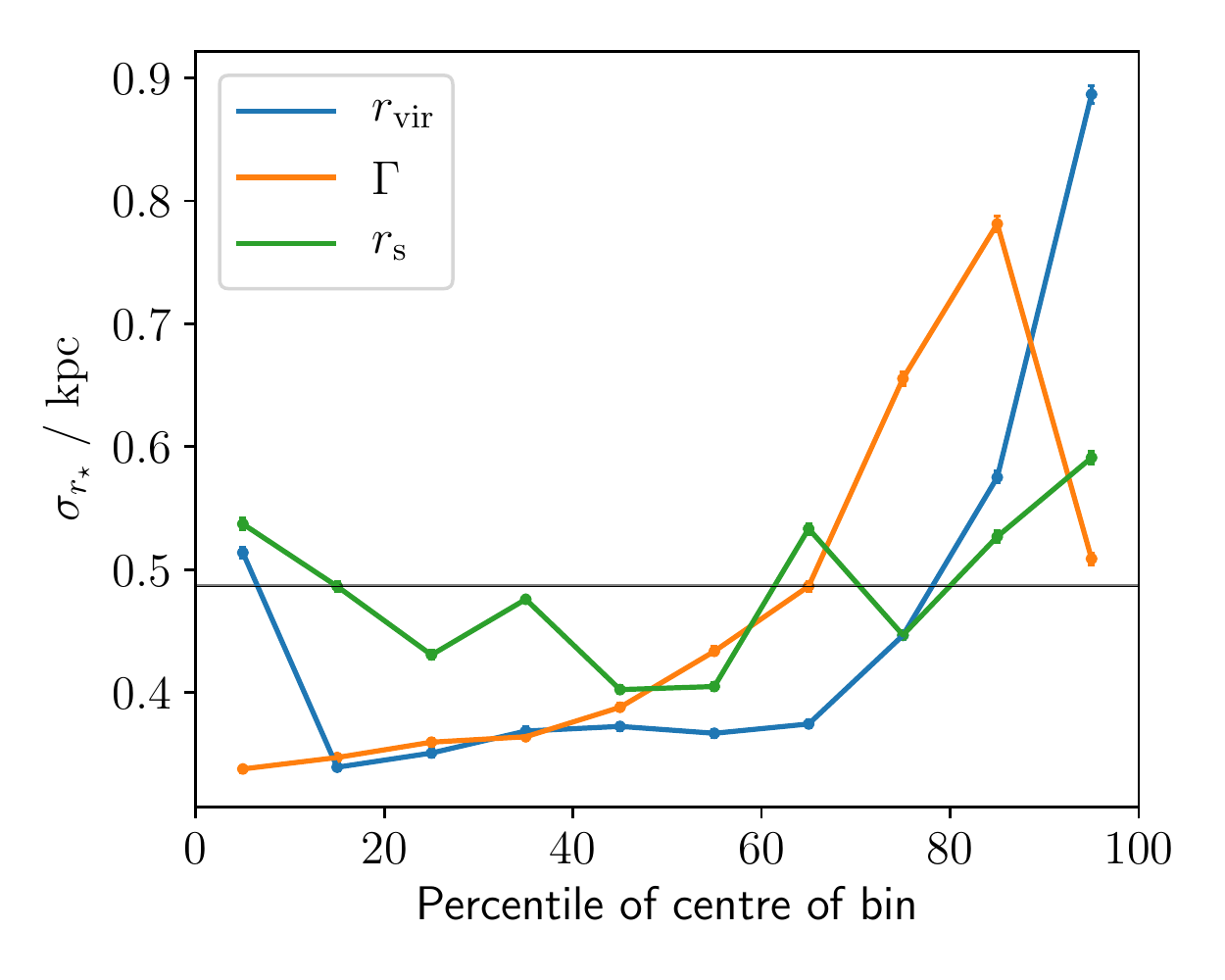}
         \label{subfig:hagn_sigma_int_binned}
     }
        \caption{
        \protect\subref{subfig:hagn_constraints_binned} Ratio of the $1\sigma$ constraints on $\Delta G / G_{\rm N}$ from the gas--star offset inference between allowing the noise parameters to vary with the halo properties vs using a single noise parameter for all galaxies. A value less than one indicates that the constraint tightens when using varying $\sigma_{r_\star}$.
        \protect\subref{subfig:hagn_sigma_int_binned} Noise parameters, $\sigma_{r_\star}$, as a function of bin number for $\lambda_{\rm c} = 4.4 {\rm \, Mpc}$, for the same bins used in \autoref{subfig:hagn_constraints_binned}. The horizontal line indicates the value obtained when a single $\sigma_{r_\star}$ is used for all galaxies.
        The constraint changes by $\lesssim 30$ per cent in all cases, weakening at large $\lambda_{\rm c}$ when we bin in $r_{\rm vir}$ due to the increase in $\sigma_{r_\star}$ for the largest halos.
        }
		\label{fig:hagn_binned}
\end{figure*}

In \autoref{subfig:hagn_constraints_binned} we compute the change in our constraints as a function of $\lambda_{\rm c}$ for 10 bins in $r_{\rm vir}$, $\Gamma$ or $r_{\rm s}$, where we compare to the constraint with a single $\sigma_{r_\star}$. Binning in any other variable produces curves similar to those of $r_{\rm s}$ and $\Gamma$. We find that allowing $\sigma_{r_\star}$ to vary can either tighten, by up to $\sim 25$ per cent, or weaken, but by no more than $\sim 30$ per cent, the constraint on $\Delta G / G_{\rm N}$. It is interesting that the constraint tightens at the smallest $\lambda_{\rm c}$ when using the more sophisticated noise model; this is the region that probes that weakest fifth force and hence sets the bound on e.g. $f_{R0}$. We have repeated this binning procedure with only 5 bins and obtain similar results, indicating that our discretisation is sufficiently fine to capture the variation of $\sigma_{r_\star}$ with these parameters.

As one would expect, the most dramatic change to our constraint occurs when we bin in the most highly correlated property, $r_{\rm vir}$. To understand the behaviour of the constraint in this case, we plot $\sigma_{r_\star}$ as a function of bin number in \autoref{subfig:hagn_sigma_int_binned}. When we bin in $r_{\rm vir}$, we find that for the majority of our bins, we obtain a smaller $\sigma_{r_\star}$ than our fiducial likelihood model, whereas the positive correlation between $r_{\rm vir}$ and the observed offset causes an increased $\sigma_{r_\star}$ for the largest halos.

\begin{figure}
	 \includegraphics[width=\columnwidth]{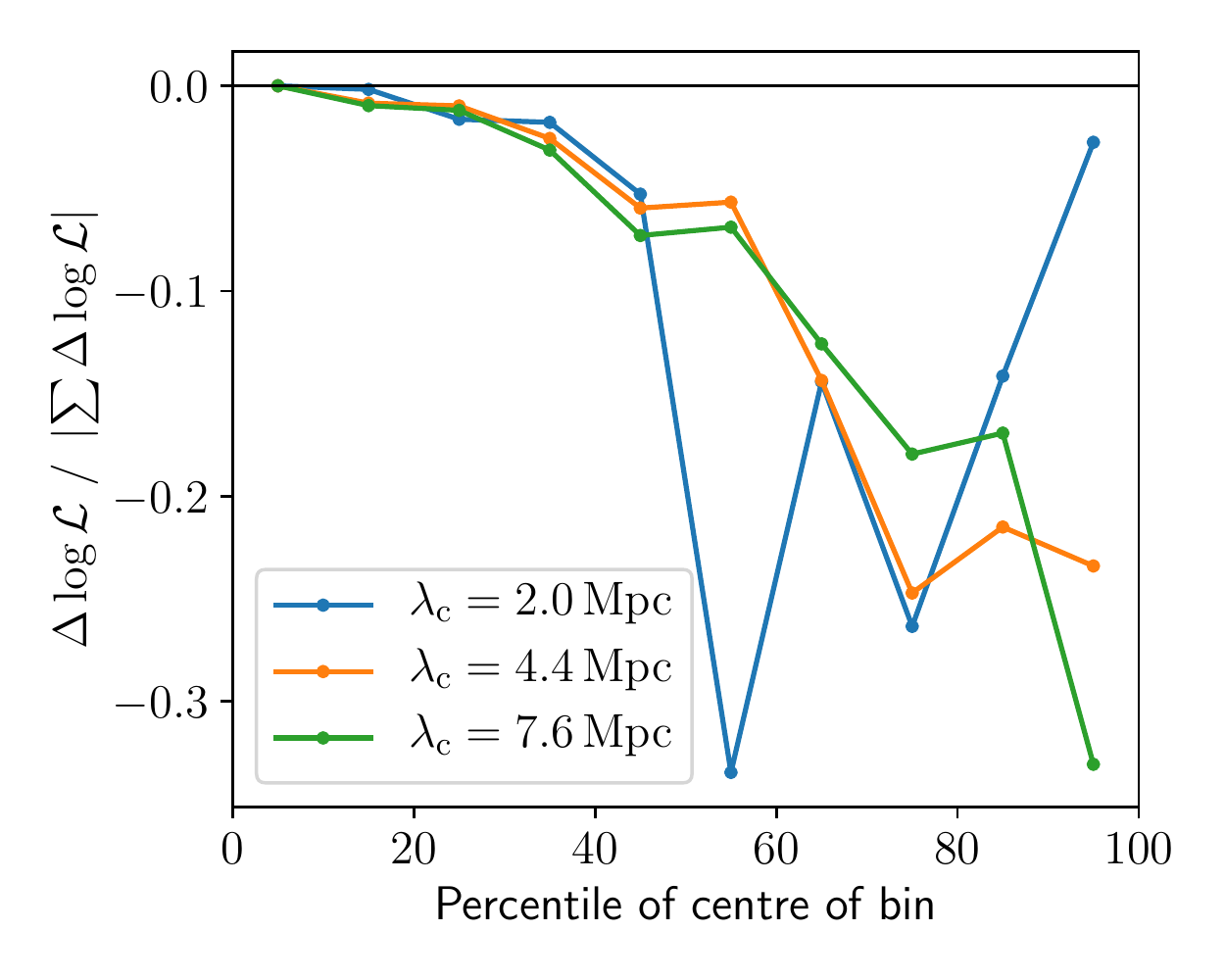}
	 \caption{\label{fig:binned_delta_loglike}The summed change in log-likelihood, $\Delta \log \mathcal{L}$, in bins of $r_{\rm vir}$ of equal size between $\Delta G / G_{\rm N} = 0$ and the $1\sigma$ constraint, for the offset analysis with a universal noise parameter. A negative $\Delta \log \mathcal{L}$ indicates a preference for $\Delta G / G_{\rm N} = 0$. We normalise by the total change of $\log \mathcal{L}$ to show the relative contribution of each bin.
	 For large values of $\lambda_{\rm c}$ the constraint is dominated by the largest $r_{\rm vir}$, but for smaller $\lambda_{\rm c}$ the galaxies with the largest halos are more likely to be screened, so that halos of intermediate $r_{\rm vir}$ are primarily responsible for the constraint.
	 }
\end{figure}

We now look at the effect each $r_{\rm vir}$ bin has on our constraint. We consider the change in the sum of the log-likelihood for all galaxies in a bin, $\Delta \log \mathcal{L}$, between the $1\sigma$ constraint on $\Delta G / G_{\rm N}$ and $\Delta G / G_{\rm N} = 0$, where we use a single $\sigma_{r_\star}$ for all bins, and set this to the maximum likelihood value. We plot the variation of $\Delta \log \mathcal{L}$ with bin number in \autoref{fig:binned_delta_loglike} for different $\lambda_{\rm c}$. For larger values of $\lambda_{\rm c}$, we see that $|\Delta \log \mathcal{L}|$ is largest for the biggest halos, i.e. our fiducial constraint is driven by the highest $r_{\rm vir}$ bins. Since these bins acquire a larger $\sigma_{r_\star}$ when we allow this to vary with $r_{\rm vir}$, the increased noise allows larger predicted signals for these galaxies. This reduces their constraining power, and hence our total constraint is weakened.
For smaller values of $\lambda_{\rm c}$ the contribution is driven by intermediate $r_{\rm vir}$, since a higher fraction of the galaxies are screened in the largest bins. Now we have the opposite case, where the noise is reduced in the bins which dominate and thus we are able to achieve tighter constraints.

As will be shown in \Cref{sec:Validity of halo density model}, the change in the constraint due to using this more complicated noise model is less than the systematic uncertainty due to the assumed halo density profile. This suggests that the simplified model used in \cite{Desmond_2018,Desmond_2018_warp,f(R)_ruled_out} is adequate given other uncertainties in the model.

By choosing just one parameter to bin in, we neglect the covariance between the parameters; it is possible that a better noise model could be constructed through a multi-dimensional binning procedure. The method we outline is easily generalisable to higher dimensions, and we have confirmed that considering two parameters simultaneously for our case study yields similar results to just using one.

So far we have assumed that the simulated signals can be measured with perfect angular resolution. As this is not the case observationally, it is important to assess how our results are affected by the addition of a realistic angular uncertainty.
To do this, we place the observer at the corner of the simulation volume to match more closely the distribution of distances of the ALFALFA survey.
We remove one galaxy which is closer to the observer than $4{\rm \, Mpc}$ (the closest any galaxy is to us in \cite{f(R)_ruled_out}).
We add a random angular displacement to the gas--star offset of each galaxy, drawn from a Gaussian of width $\sigma_{\rm obs}$. We consider two cases: 1) $\sigma_{\rm obs}=18\arcsec$ to mimic the ALFALFA survey and thus the inference of \cite{Desmond_2018,f(R)_ruled_out}, and 2) $\sigma_{\rm obs}=0 \farcs 1$, as will be achievable with the `mid’ configuration of SKA1 \cite{Santos_2015,Yahya_2015}. Now  we must fit for both an angular, $\sigma_{\rm obs}$, and intrinsic, $\sigma_{\rm int}$, noise component, such that the appropriate $\sigma_{r_\star}$ for galaxy $i$ at distance $d_{i}$ is 
\begin{equation}
    \sigma_{{r_\star}, i} = \sqrt{\sigma_{\rm int}^2 + \left( \sigma_{\rm obs} d_{i} \right)^2}.
\end{equation}

\begin{figure}
	 \includegraphics[width=\columnwidth]{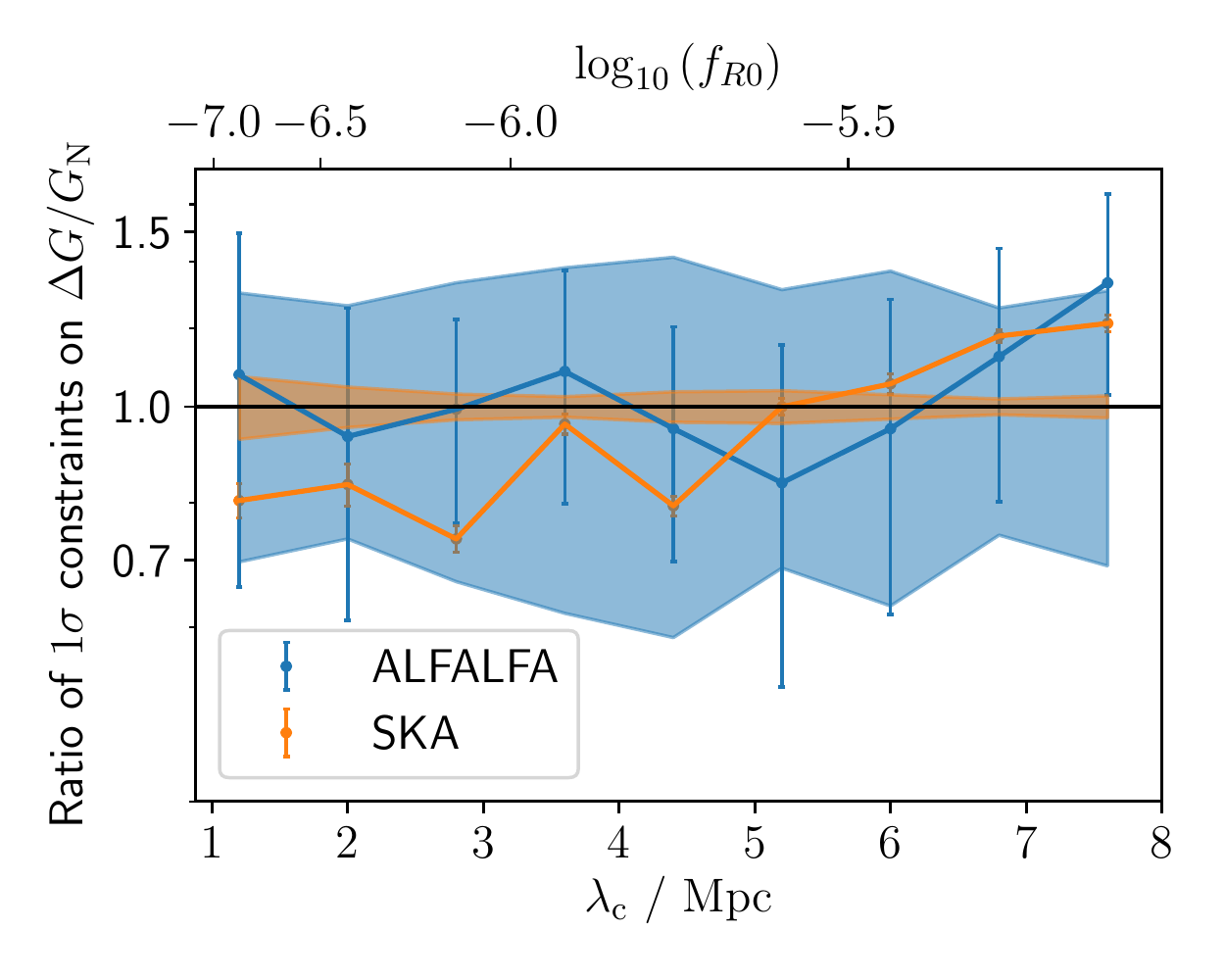}
	 \caption{\label{fig:hagn_constraints_binned_added} Same as \autoref{subfig:hagn_constraints_binned}, but here we add an additional random angular offset to each galaxy at either the ALFALFA ($18\arcsec$) or SKA ($0\farcs1$) resolution. Our binned noise model has a universal angular uncertainty combined with a different intrinsic contribution for each of 5 bins in $r_{\rm vir}$, where each bin contains the same number of galaxies. We run the un-binned and binned analyses 15 times each, and the standard deviation of the constraint between these is given by the shaded region and error bars respectively.}
\end{figure}

We now re-run the inference with this additional, random angular offset 15 times for each $\lambda_{\rm c}$, fitting in each case for a universal $\Delta G / G_{\rm N}$ and $\sigma_{\rm obs}$, and either a universal $\sigma_{\rm int}$ or a different $\sigma_{\rm int}$ for each of 5 bins in $r_{\rm vir}$. The change in the constraint from using a $r_{\rm vir}$-dependent $\sigma_{\rm int}$ is shown in \autoref{fig:hagn_constraints_binned_added}. We see that for both the ALFALFA and SKA resolution, the mean constraint changes by $\lesssim 30$ per cent for all $\lambda_{\rm c}$.

It is unsurprising that the ALFALFA-like constraints are insensitive to whether we bin $\sigma_{\rm int}$ or not, with the $1\sigma$ regions of the uncertainty on the constraint overlapping for all $\lambda_{\rm c}$. An $18 \arcsec$ angular offset at the mean distance from the observer ($128 {\rm \, Mpc}$) corresponds to $11 {\rm \, kpc}$, so the angular offset dominates the intrinsic contribution of $\sim 0.5 {\rm \, kpc}$. One would expect that changing the model of the subdominant contribution to the noise would have little impact on the constraint, as indeed we find.

For the SKA resolution we are in the opposite regime: a $1\sigma$ angular offset now corresponds to only $60 {\rm \, pc}$ at the mean distance to a galaxy, and thus our results are dominated by the intrinsic component. This results in a smaller sample variance than with the ALFALFA resolution, and a practically identical variation of the change of the constraint with $\lambda_{\rm c}$ as in \autoref{subfig:hagn_constraints_binned}. Now the $1 \sigma$ uncertainties on the constraints from multiple runs do not overlap, but again the constraints change by less than 30 per cent for all $\lambda_{\rm c}$.

The uncertainty on the constraint is actually larger than the variation across these realisations, as will be discussed in \Cref{sec:Validity of halo density model}. Given this, and that the intrinsic contribution to the offset here is expected to be an upper limit set by the simulation's resolution, we conclude that the assumption of uncorrelated Gaussian noise is justified given the other potential systematic uncertainties in the inference at both the ALFALFA and SKA resolution. This is to be expected, as we previously found this to be true even in the absence of an angular uncertainty.

\section{Validity of halo density model}
\label{sec:Validity of halo density model}

\begin{figure}
	 \includegraphics[width=\columnwidth]{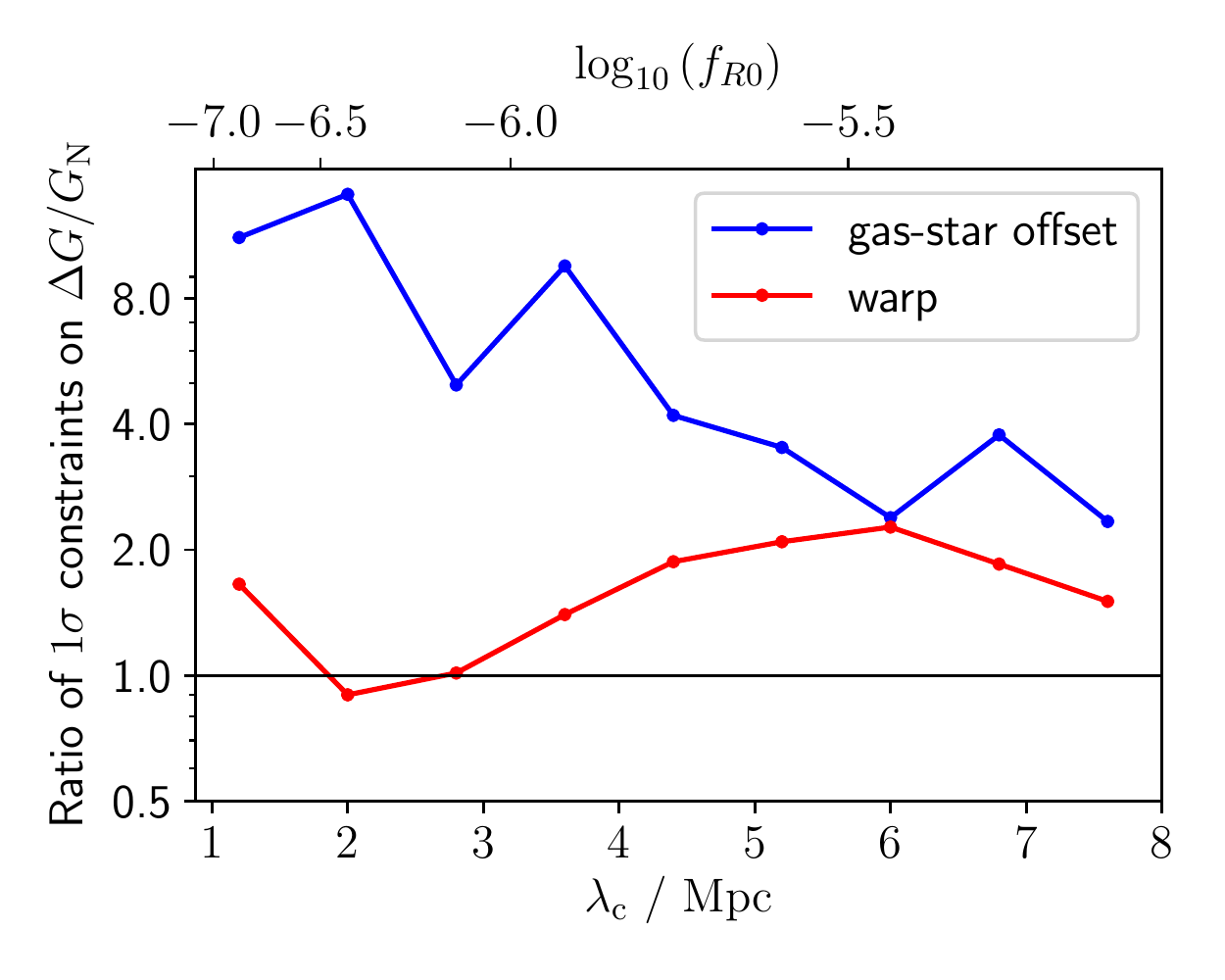}
	 \caption{\label{fig:hagn_constraints_gennfw}Ratio of the $1\sigma$ constraints on $\Delta G / G_{\rm N}$ between assuming an inner power law of slope $\beta=0.5$ inside the scale radius of a NFW profile, and using a more general NFW-like profile, where we explicitly fit for the inner power law slope. A value greater than one indicates a weaker constraint when using the more general density profile.}
\end{figure}

We now wish to see how the assumed form of the halo density profile affects our constraints, a test afforded by the full dark matter distribution available in the simulation. We consider the change in the constraint from assuming $\beta = 0.5$ with the NFW properties to using $\Gamma$ as the inner power law index alongside the parameters from the more general NFW-like profile.
To isolate the impact of the halo density profile from the number of galaxies used to make the constraint, when making this comparison we only use galaxies which have a non-zero template signal with both parametrisations. The resulting changes in the constraints are plotted in \autoref{fig:hagn_constraints_gennfw}. From this we see that, as in \cite{f(R)_ruled_out}, the constraints from the warp analysis vary by less than a factor of $\sim 2$ for most $\lambda_{\rm c}$, although we do note that the constraint tends to be weaker when using the more general density profile.
The gas--star offset analysis appears to be more strongly affected by the assumed density model, with constraints up to an order of magnitude weaker with the more general density profile.

The conclusion to draw from this analysis is conditioned on the reliability of the halo density profiles in Horizon-AGN. The inner slope of the halo is determined by the balance between adiabatic contraction \cite{Blumenthal, Gnedin}, which steepens the profile, and stellar or AGN feedback, which promotes core formation \cite{Pontzen_Governato, DP_CuspCore}. While redshift-zero halos in Horizon-AGN are found to be steeper than NFW \cite{Peirani_2017}, most observational evidence favours shallower profiles (e.g.~\cite{Oh,Salucci}), a manifestation of the cusp-core problem of $\Lambda$CDM \cite{Cusp_core}. It is unclear how alternative feedback prescriptions would alter the constraints in \autoref{fig:hagn_constraints_gennfw}, although we note that by considering only halos with non-zero template signals we have removed all those with $\Gamma < -1$ (\Cref{sec:Modelling the halo restoring force}).

Due to this issue, we cannot a draw a definitive conclusion about the validity of the halo density model used in \cite{f(R)_ruled_out}. We can however say that, for the Horizon-AGN simulation, variations in the halo density model can result in a weakening of the constraint of $\Delta G / G_{\rm N}$ from gas--star offsets by up to a factor $\sim 10$, and galaxy warps by up to a factor $\sim 2$. Although this is consistent with the discussion of systematic uncertainties in \cite{f(R)_ruled_out}, different simulations or different halo density profiles could lead to different conclusions. It is therefore important for future work to investigate other cosmological hydrodynamical simulations with different implementations of galaxy formation physics, to determine whether this result is robust to plausible variations in the dark matter distribution.

For this specific example, we note that it is not the individual constraints that are important but rather the combined constraint of the two analyses. From figure 1 of \cite{f(R)_ruled_out}, we see that the combined constraint on $\Delta G / G_{\rm N}$ would hardly change if we removed the gas--star offset analysis and only used galaxy warps. The sensitivity of the gas--star offset inference to the halo density profile is therefore of little consequence to the conclusions of that work, which are driven mainly by the warp analysis.

\section{Discussion and Conclusions}
\label{sec:Conclusions}

Galactic scale tests are capable of providing powerful constraints on new fundamental physics across a range of previously under-explored environments. Their drawback is that one must accurately model the messy astrophysical regime to capture fully, and break degeneracies with, baryonic effects that are less important in more traditional analyses.

We propose three main tests to gauge the robustness of a given model to baryonic physics:
\begin{enumerate}
    \item Determine whether there are any unaccounted-for correlations between the target observable (signal) and galactic properties due to baryonic physics (noise). To do this, we ask whether the signal in a cosmological hydrodynamical simulation without new physics can be predicted from the parameters relevant to the new physics model. Since the functional form of any such correlation is unknown a priori, this is best addressed in a machine-learning context.
    \item If one or more parameter is found to correlate with the simulated signal, then the impact of this on the new physics constraint must be established. We do this by explicitly constructing a more sophisticated noise model and repeating the inference. If the constraint changes by less than some specified tolerance then it can be concluded that the simplified noise model is satisfactory; otherwise it should be replaced with the more complex one.
    \item Use the extra information available in the simulation compared to observations to assess the adequacy of the modelling of unobservable properties. With the simulated data one can compare the constraints obtained using the ``true'' vs model parameters to quantify the model's suitability, and improve it if necessary.
\end{enumerate}

As a case study, we investigate the morphological signatures used in \cite{f(R)_ruled_out} to constrain thin-shell-screened fifth forces and rule out astrophysically relevant Hu-Sawicki $f(R)$ theories, namely offsets between the stellar and gaseous components of a galaxy and warping of the stellar disk. We work in the context of the Horizon-AGN simulation, performed in $\Lambda$CDM and hence with no fifth force. We use a machine-learning feature importance analysis to study the correlations in the simulation between the morphological signals and the parameters used to predict them in the context of modified gravity. We find that the degree of `U'-shaped warping of the stellar disk is independent of these parameters, justifying the assumption of uncorrelated random Gaussian noise used in \cite{Desmond_2018_warp,f(R)_ruled_out}. For the gas--star offset case, we find a positive correlation between the signal and the virial radius of the host halo, $r_{\rm vir}$. To assess the impact of this previously unaccounted-for complication, we allow the width of the Gaussian noise model to vary with $r_{\rm vir}$ by binning the galaxies in $r_{\rm vir}$ and using a different width in each bin. This changes the fifth force constraint by $\lesssim 30$ per cent, which is less than the systematic uncertainty from the assumed halo density profile. This again justifies the assumption of random noise. We find that the gas--star offset inference is more sensitive to the assumed density profile than the warp analysis, and we identify this as the largest source of systematic uncertainty in the inference of \cite{f(R)_ruled_out}. It could be reduced in the future using dynamical information on the central regions of the test galaxies.

We anticipate the methods developed here to prove useful for validating future forward models of galactic signals used to constrain fundamental physics. In this work, we have only considered the implementation of galaxy formation physics in the Horizon-AGN simulation. Other simulations have different sub-grid models, calibration methods, hydrodynamical and feedback schemes and resolutions, and any analysis must be robust to these differences. It is known that small-scale predictions for the matter power spectrum are sensitive to these effects \cite{Chisari_2019}, so it is natural to suspect a similar sensitivity here. Future studies should therefore apply the methods we outline to different simulations and different physics tests to assess the accuracy of constraints derived from the galactic regime.

\acknowledgements
{
We thank Eliza Dickie for early contributions to this work.

We would like to thank the Horizon-AGN collaboration for allowing us to use the simulation data, and particularly Stephane Rouberol for smoothly running the Horizon Cluster hosted by the Institut d'Astrophysique de Paris where most of the processing of the raw simulation data was performed. Some of the numerical work also made use of the DiRAC Data Intensive service at Leicester, operated by the University of Leicester IT Services, which forms part of the STFC DiRAC HPC Facility (\url{www.dirac.ac.uk}). The equipment was funded by BEIS capital funding via STFC capital grants ST/K000373/1 and ST/R002363/1 and STFC DiRAC Operations grant ST/R001014/1. DiRAC is part of the National e-Infrastructure.

DJB is supported by STFC and Oriel College, Oxford. HD is supported by St John's College, Oxford. PGF is supported by the STFC. HD and PGF acknowledge financial support from ERC Grant No 693024 and the Beecroft Trust. 
}

\bibliographystyle{apsrev4-1}
\bibliography{references}

\end{document}